\documentclass[journal]{IEEEtran}

\usepackage{graphicx}
\usepackage{amssymb}
\usepackage{adjustbox}

\usepackage{multirow}
\usepackage{xcolor,colortbl}

\usepackage{amsthm}

\usepackage{booktabs}

\usepackage{amsmath}

\usepackage[numbers]{natbib}
\newcommand*\reviewcolor{\color{black}}

\begin{document}

\title{End-to-End Prediction of Parcel Delivery Time with Deep Learning for Smart-City Applications}

\author{Arthur~Cruz~de~Araujo,~Ali~Etemad
\\\textit{Department of Electrical and Computer Engineering}\\\textit{Queen's University}\\\textit{Kingston, Canada}\\\{18acda, ali.etemad\}@queensu.ca

}
    

\maketitle

\begin{abstract} 
The acquisition of massive data on parcel delivery motivates postal operators to foster the development of predictive systems to improve customer service. Predicting delivery times successive to being shipped out of the final depot, referred to as \textit{last-mile} prediction, deals with complicating factors such as traffic, drivers' behaviors, and weather. This work studies the use of deep learning for solving a real-world case of last-mile parcel delivery time prediction. We present our solution under the IoT paradigm and discuss its feasibility on a cloud-based architecture as a smart city application. We focus on a large-scale parcel dataset provided by Canada Post, covering the Greater Toronto Area (GTA). We utilize an origin-destination (OD) formulation, in which routes are not available, but only the start and end delivery points. We investigate three categories of convolutional-based neural networks and assess their performances on the task. We further demonstrate how our modeling outperforms several baselines, from classical machine learning models to referenced OD solutions. We perform a thorough error analysis across the data and visualize the deep features learned to better understand the model behavior, making interesting remarks on data predictability. Our work provides an end-to-end neural pipeline that leverages parcel OD data as well as weather to accurately predict delivery durations. We believe that our system has the potential not only to improve user experience by better modeling their anticipation but also to aid last-mile postal logistics as a whole.
\end{abstract}

\begin{IEEEkeywords}
last-mile, parcel delivery, origin-destination, predictive modeling, deep learning.
\end{IEEEkeywords}

\section{Introduction}

The massive accumulation of data is rooted in the recent advances in the ubiquity of instrumentation, especially in urban environments and everyday products. Sensor systems such as Global Positioning System (GPS) now pervade day-to-day life and are at the core of the technology that makes smart cities possible, an evolving concept that gravitates towards the use of technology to enhance the quality of life of citizens, while attaining a sustainable use of natural resources \cite{ismagilova2019smart}.

Among the various aspects of smart cities, this paper focuses on smart transportation. Tracking time is of most interest in private and public transport services, for vehicles like buses, taxis, and trucks. Estimating time has become even more urgent with traffic continuously worsening and urban mobility getting increasingly complex. For instance, congestion rose by $36\%$ in Los Angeles and $30\%$ in New York from 2010 to 2016, aggravating health issues related to accidents and air pollution \cite{mckinsey_urban_mobility}. With rapid progress on smart vehicles, GPS data can improve travel time estimates, aiding traffic management, urban planning, and fuel consumption optimization.

The huge growth in retailing and the ever-increasing draw of customers to e-commerce point to the need for a reformulation of delivery logistics. For retail companies, providing accurate delivery time estimates is of major importance, for both customer satisfaction and internal organization, since it allows, for example, for better routing and drivers' scheduling. Also, the existence of large competitors with fast solutions such as next-day deliveries pushes retailers to direct resources to improve their logistics to preserve their market share. Moreover, an accurate \textit{estimate of delivery time} has shown to be of considerable importance to the customer.

We tackle the problem of last-mile parcel delivery time estimation. A parcel is scanned at many depots before reaching its final destination. The qualifier \textit{last-mile} denotes the last segment of that journey, from the moment the parcel is sent out for delivery to the moment it gets delivered. Formally, last-mile logistics refers to the last stretch of a business-to-consumer parcel delivery service, from the order penetration point to the consignee’s preferred destination point \cite{lim2018consumer}. The final segments of the delivery can account for $13-75\%$ of the overall supply chain cost \cite{gevaers2009characteristics}. When it comes to the impact of poor estimates on customer satisfaction, the last-mile is the most important segment to be modeled given the closeness to the delivery time and the anticipation by the customer. Furthermore, by providing accurate delivery times at that stage, the retailer can properly manage the expectations of the customer.

Most solutions on travel time estimation require information on the traveled route \cite{Chen_2004, gal2017traveling, wang2018learning, wang2018will, zhang2018deeptravel}. Although applicable in the real world, they introduce uncertainty when requiring a pre-planned route for prediction, which in practice may be impossible. This results in high demand for solutions that only depend on the start and end points of the trip, referred to as Origin-Destination Travel Time Estimation (OD-TTE). OD-based methods generally make use of large datasets for modeling spatio-temporal relationships \cite{guo2012discovering, zhan2013urban, alexander2015origin, wang_simple, Bertsimas_2019, 2017arXiv171004350J, li2018multi, deep_net_travel_2019}. Some of them tackle delivery time prediction based on the geographical coordinates of the origin and destination, and the week, day, and time when the delivery started \cite{alexander2015origin, wang_simple, deep_net_travel_2019}.

Despite the availability of a range of solutions and learning techniques for predicting delivery times, this problem remains challenging due to a number of open problems:
\begin{itemize}
    \item Delivery times are strongly reliant on human driving, which is a highly unpredictable phenomenon, despite the existence of standards and protocols defined by delivery companies in general that might not fully hold in practice. 
    \item Hard-to-predict factors such as weather conditions and traffic variations can have a high impact on delivery time.
    \item Company-specific logistics involving drivers' scheduling and route planning will also affect travel time estimation.
\end{itemize}

In this paper, we investigate a series of deep learning models for tackling the problem of parcel delivery time estimation, formulated as an OD-TTE problem. We explore instances from three different categories of convolutional-based neural networks and evaluate their performances on the task. Namely, we experiment with VGG (Visual Geometry Group) architectures \cite{simonyan2014deep} as a very popular type of convolutional neural network (CNN) used in a variety of domains; residual neural networks (ResNet) \cite{He_2016_CVPR}, given their strong ability to learn effective representations while avoiding the vanishing gradient problem as a result of introducing the skip connections; and squeeze-and-excitation (SE) operators \cite{Hu_2018_CVPR}, which attempts to better capture channel interdependencies in the data through feature recalibration. We thoroughly analyze these categories of neural networks, and a number of other benchmarks, to identify the best approach capable of accurately predicting last-mile parcel delivery times. While our approach, in general, makes no attempt at directly predicting traffic or human behavior patterns, it provides a direct end-to-end neural pipeline tailored to leverage parcel data information and return accurate predictions that might be further utilized to improve last-mile logistics. We validate the system on a real-world last-mile parcel delivery dataset, collected and provided by Canada Post \cite{canadapost}, the main postal operator in Canada. The dataset comprises last-mile information from parcels delivered in the Greater Toronto Area (GTA) within the period from January 2017 to June of the same year, including spatial and temporal data, as well as weather measurements. We compare all of our deep models against a number of benchmarks, including classical machine learning approaches and existing OD solutions from the literature, and we demonstrate the effectiveness of our system in comparison to the baselines. {\reviewcolor The large size and complexity of the dataset and recent success of deep learning solutions in modeling such complex data in other fields as well as travel-time estimation \cite{wang2018will,zhang2018deeptravel,2017arXiv171004350J,li2018multi,deep_net_travel_2019}, motivates our choice of deep learning for the solution. The performance of the proposed system in outperforming classical modeling techniques further validates our solution for this particular task and dataset.}

Our contributions in this paper are summarized as follows:
(1) we propose an end-to-end system capable of parcel delivery time prediction under the IoT paradigm for smart city applications. To this end, we study the application of a series of deep CNNs towards solving this problem, relying solely on start and end points, with no knowledge of the route taken. Our solution leverages spatio-temporal information, while also taking weather into account, using daily temperature, rain and snow precipitation, and the amount of snow in the ground;
(2) we validate our solution on a very large \textit{real-world} dataset on parcel delivery in Canada's largest metropolitan area (GTA), provided by Canada's main postal operator Canada Post;
(3) our end-to-end solutions achieve competitive results with accurate performance, outperforming the benchmark techniques as well as a number of methods in the related work.

\section{Related Work}

\subsection{Last-Mile Parcel Delivery}

According to \cite{olsson2019framework}, last-mile logistics can be divided into five components: logistics, distribution, fulfillment, transport, and delivery. Much of the last-mile research proposes new formulations for logistics or attempts to improve specific aspects of it. For example, the work in \cite{punakivi2001solving} compares the usage of delivery boxes to reception boxes for deliveries, concluding that the unattended reception of parcels can reduce up to $60\%$ of delivery costs. In \cite{aurambout2019last}, the potential of drones for last-mile delivery and the accessibility to such service is investigated. In \cite{wanganoo2020preparing}, a conceptual last-mile framework is proposed integrating technologies such as GPS and IoT to enhance the delivery operator visibility through a smartphone. Furthermore, many works have proposed solutions to improve last-mile deliveries by using car trip sharing \cite{Crowdphysics, Wang_Zhang_Liu_Shen_Lee_2016, crowddeliver_2017, Ridesharing_2018,  wang2019demystifying, Car4Pac}.

Although a lot can be found on last-mile logistics and delivery, sources become scarce when it comes to the specific application of travel time estimation. A recent work \cite{wu2019deepeta} proposes DeepETA, a spatial-temporal sequential neural network model for estimating travel (arrival) time for parcel delivery, formulated with multiple destinations and taking into account the latest route sequence by knowing the order in which the parcels were delivered. The model overcomes state-of-the-art methods on their real logistics dataset. However, this approach relies on the availability of the delivery routes, which renders it incompatible with our work's OD-based formulation.

\subsection{Travel Time Estimation}
Travel time estimation is an area concerned with estimating travel durations from data. While it encompasses last-mile delivery, most works in the area have focused on other applications such as taxi trip durations \cite{wang2018will, zhang2018deeptravel, guo2012discovering}. The literature on travel time estimation is often subdivided into two categories: solutions that depend on knowing the traveled routes, hence called \textit{route-based} methods, and those that only require the start and end points of the trip, being referred to as \textit{Origin-Destination Travel Time Estimation} (OD-TTE) \cite{li2018multi}. Recent effort has been made towards OD research, encouraged by the fact that in many cases, due to either privacy concerns, tracking costs, or the lack of ability to fully pre-plan the route, only limited information is available \cite{wang_simple}.

\subsubsection{Route-based Travel Time Estimation}
In \cite{Chen_2004} a neural network was used to predict bus travel times. Given hour-of-day, day-of-week, and weather conditions (precipitation), the network would estimate the travel time for each route segment, further adjusting the prediction using Kalman filtering and up-to-the-minute bus locations. Similarly, the work in \cite{gal2017traveling} estimates the duration of a bus ride based on a scheduled route and a source-destination pair. In \cite{wang2018learning}, travel time was predicted based on floating-car data, by combining linear models, deep, and recurrent neural networks to explore road segment sequences. In \cite{wang2018will}, DeepTTE was proposed, an end-to-end deep learning framework that predicts travel time for an entire path as a whole. Associating convolutional and recurrent stages, the approach experimented on two trajectory datasets and outperformed state-of-the-art methods. Similarly, DEEPTRAVEL \cite{zhang2018deeptravel} estimates the travel time of a whole trip directly, given a query path. The paper proposed a feature extraction structure to effectively capture spatial, temporal, and traffic dynamics, allowing for accurate estimations.

\subsubsection{OD Travel Time Estimation}
An approach was presented in \cite{guo2012discovering} to discover spatio-temporal patterns in OD displacements. Spatial clustering of coordinates was used to identify meaningful places, and flow measures were extracted from the clusters to understand the spatial and temporal trends of movements. In \cite{zhan2013urban}, taxi OD data was used for estimating the hourly average of travel times for urban links. For each trip, possible paths were inferred and the link travel times were estimated. The work in \cite{alexander2015origin} estimated daily OD trips from triangulated mobile phone records, converting the phone records into clustered locations where users engaged in activities for an observed duration. Trips were then constructed for each user between two consecutive observations in a day and scaled up using census data. 
Rather than using routes, OD-TTE was directly tackled in \cite{wang_simple}, where a \textit{Simple Baseline to Travel Time Estimation} (SB-TTE) was proposed. The work used large-scale trip data to estimate travel time between two points based on origin, destination, and \textit{actual distance traveled}. \textbf{It should be noted, however, that actual traveled distance is not available in real-world settings prior to the travel taking place and without a pre-planned route.} In \cite{Bertsimas_2019}, insights were borrowed from network optimization to demonstrate that it is possible to extract accurate travel time estimations from a large OD dataset to reconstruct the traffic patterns in a whole city. In \cite{2017arXiv171004350J}, a model called Spatio-Temporal Neural Network (ST-NN) is proposed for OD travel time estimation, which jointly predicts trip durations and traveled distances. In \cite{li2018multi}, a MUlti-task Representation learning model for Arrival Time estimation (MURAT) was proposed, a solution claimed to produce meaningful data representations leveraging the underlying road network and spatio-temporal priors. Although an OD-based approach, MURAT requires route information for training purposes. Finally, in \cite{deep_net_travel_2019}, we explored an NYC taxi trip dataset in the context of OD-TTE. In that work, deep neural networks were used for estimation travel times, outperforming a number of baseline techniques.

Table \ref{related_work_summary} summarizes the related work, organized according to problem type, that is, between route-based and route-free methods. The table also specifies what type of datasets were used, for instance, bus records, taxis, etc. It also lists what information is used as input data. Lastly, Table \ref{related_work_summary} outlines the prediction method of each work so as to better understand the variety of tools that have been used in this problem domain.

\newcommand{\colrot}[3]{
\parbox[t]{2mm}{\multirow{-#1}{*}{\cellcolor[#2]{0.9}\rotatebox[origin=c]{90}{#3}}}
}
\newcommand{\customcite}[1]{
\multirow{1}{*}{\citet{#1}} & \multirow{1}{*}{\citeyear{#1}}
}
\newcommand{\custpbox}[2]{
\multirow{#1}{*}{\parbox{5cm}{#2}}
}
\newcommand{\custpboxb}[2]{
\multirow{#1}{*}{\parbox{3cm}{#2}}
}
\newcommand\Tstrut{\rule{0pt}{2ex}}         
\newcommand\Bstrut{\rule[-0.9ex]{0pt}{0pt}} 
\newcommand{\TBstrut}{\Tstrut\Bstrut}       

\begin{table*}[!t]
\caption{Related Work Summary}
\label{related_work_summary}
\scriptsize
\centering
\begin{adjustbox}{width=\textwidth}
\begin{tabular}{l|l|l|l|l|l|l}
\hline
& Reference & Year & Data Type & Location & Input Data & Prediction Method\\
\hline
\hline
\cellcolor[gray]{0.9} & \multirow{1}{*}{\citet{Chen_2004}} & \multirow{1}{*}{\citeyear{Chen_2004}}  & \multirow{1}{*}{Bus}          & \multirow{1}{*}{New Jersey}       & \custpbox{1}{Day-of-week, timestamp, segment, precipitation (rain/snowfall)}                                &   \multirow{1}{*}{Neural Networks; Kalman filtering}      \Tstrut  \\
\cellcolor[gray]{0.9} &                 &                   &                               &                                   &                                                                                                               &                                                           \Bstrut  \\ 
\cline{2-7}
\cellcolor[gray]{0.9} & \customcite{gal2017traveling}  & \multirow{1}{*}{Bus}          & \multirow{1}{*}{Dublin}           & \custpbox{2}{Day-of-week, timestamp, travel time of the last bus using the segment, and time passed since then}     &   \multirow{1}{*}{Snapshot principle; decision trees}     \Tstrut  \\
\cellcolor[gray]{0.9} &                 &                   &                               &                                   &                                                                                                                       &                                                                    \\
\cellcolor[gray]{0.9} &                 &                   &                               &                                   &                                                                                                                       &                                                           \Bstrut  \\ 
\cline{2-7}
\cellcolor[gray]{0.9} & \customcite{wang2018learning}  & \multirow{1}{*}{Floating-car} & \multirow{1}{*}{Beijing}          & \custpbox{3}{GPS, time period in a year, a month and a day, holiday indicator, rush hour indicator, weather (no details given), driver profile}   &   \multirow{3}{*}{Deep neural networks and LSTM}  \Tstrut  \\
\cellcolor[gray]{0.9} &                 &                   &                               &                                   &                                                                                                                                                   &                                                            \\
\cellcolor[gray]{0.9} &                 &                   &                               &                                   &                                                                                                                                                   &                                                   \Bstrut  \\
\cline{2-7}
\cellcolor[gray]{0.9} & \customcite{wang2018will}  & \multirow{1}{*}{Taxi}         & \multirow{1}{*}{Chengdu/Beijing}  & \custpbox{3}{GPS, distance of path, day-of-week, time-slot of travel start, weather condition (rainy, sunny, windy), driver ID}                   &   \multirow{1}{*}{LSTM}                           \Tstrut  \\
\cellcolor[gray]{0.9} &                 &                   &                               &                                   &                                                                                                                                                   &                                                            \\ 
\cellcolor[gray]{0.9} &                 &                   &                               &                                   &                                                                                                                                                   &                                                   \Bstrut  \\ 
\cline{2-7}
\cellcolor[gray]{0.9} & \customcite{zhang2018deeptravel}  & \multirow{1}{*}{Taxi}         & \multirow{1}{*}{Porto/Shanghai}   & \custpbox{1}{GPS, timestamp, driving state}                                                                                                       &   \multirow{1}{*}{BiLSTM}                                  \TBstrut \\
\cline{2-7}
\colrot{13}{gray}{Route-based} & \customcite{wu2019deepeta}  & \multirow{1}{*}{Parcel}      & \multirow{1}{*}{Beijing}          & \custpbox{1}{Latest route, GPS, date, deliver status}         &   \multirow{1}{*}{BiLSTM}                                  \TBstrut \\
\hline
\cellcolor[gray]{0.9} & \customcite{guo2012discovering}     & \multirow{1}{*}{Taxi}         & \multirow{1}{*}{Shenzhen}         & \custpbox{1}{GPS, timestamp}                                  &   \multirow{1}{*}{Clustering; statistical summaries}      \TBstrut \\   
\cline{2-7}
\cellcolor[gray]{0.9} & \customcite{zhan2013urban}     & \multirow{1}{*}{Taxi}         & Small area                        & \custpbox{1}{GPS, road network, hourly intervals}             &   Multinomial logistic for path                   \Tstrut \\
\cellcolor[gray]{0.9} &                 &                               &                               & in Midtown                        &                                                               &   probability; then, least square              \TBstrut \\ 
\cellcolor[gray]{0.9} &                 &                               &                               & Manhattan                         &                                                               &   optimization for link travel time              \Bstrut \\ 
\cline{2-7}
\cellcolor[gray]{0.9} & \customcite{alexander2015origin}     & Mobile                        & Boston                            & \custpbox{3}{Call detail records (CDRs) with time-stamped GPS coordinates of anonymized customers; day-of-week, hour-of-day}          & Rule-based user activity inference;               \Tstrut \\
\cellcolor[gray]{0.9} &                 &                               & phone                         &                                   &                                                                                                                                       & probabilistic estimation of              \\ 
\cellcolor[gray]{0.9} &                 &                               & records                       &                                   &                                                                                                                                       & departure times                                                   \Bstrut \\
\cline{2-7}
\cellcolor[gray]{0.9} & \customcite{wang_simple}     & \multirow{1}{*}{Taxi}         & NYC/Shanghai                      & \custpbox{2}{GPS, travel distance, departure time (weekly-basis and absolute)}    & Neighbor-based estimation     \Tstrut \\
\cellcolor[gray]{0.9} &                 &                               & records                       &                                   &                                                                                   &                               \Bstrut \\
\cline{2-7}
\cellcolor[gray]{0.9} & \customcite{2017arXiv171004350J}     & \multirow{1}{*}{Taxi}         & NYC                               & \custpbox{1}{GPS, travel distance, departure timestamp}                           & Multi-Layer Perceptron (MLP)               \TBstrut \\ 
\cline{2-7}
\cellcolor[gray]{0.9} & \customcite{li2018multi}     & \multirow{1}{*}{Taxi/Didi}    & Beijing/NYC                       & \custpbox{3}{GPS, travel distance, departure day-of-week and hour-of-day, and during training the trip path is used}      & ResNet-based neural network   \Tstrut \\
\cellcolor[gray]{0.9} &                 &                               &                               &                                   &                                                                                                                           &                                       \\ 
\cellcolor[gray]{0.9} &                 &                               &                               &                                   &                                                                                                                           &                               \Bstrut \\ 
\cline{2-7}
\cellcolor[gray]{0.9} & \custpboxb{2}{\citet{deep_net_travel_2019}} & \multirow{1}{*}{\citeyear{deep_net_travel_2019}}      & \multirow{1}{*}{Taxi}             & NYC                               & \custpbox{1}{GPS, day-of-week and hour-of-day taken at departure}                                                    & Multi-Layer Perceptron (MLP)          \Tstrut \\
\cellcolor[gray]{0.9} &                 &                               &                               &                                   &                                                                                                               &                               \Bstrut \\ 
\cline{2-7}
\colrot{16}{gray}{OD-based} & \customcite{Bertsimas_2019}     & \multirow{1}{*}{Taxi}         & NYC                               & \custpbox{1}{GPS, date and time}                                                    & Shortest-path convex optimization          \TBstrut \\
\hline
\end{tabular}
\end{adjustbox}
\end{table*}

\section{System Architecture}

\subsection{Smart City Contextualization}

Smart transportation is one of the key services that define a smart city, being indeed an integral part of any smart city initiative. Smart transport systems are meant to improve the safety, security, efficiency, and environmental friendliness of transport systems, by combining Intelligent Transportation Systems (ITS) with other vehicular technologies such as autonomous vehicles \cite{song2017smart}. The deployment of smart transportation is enabled by technologies such as GPS and the Internet of Things (IoT), a communication paradigm according to which objects of everyday life will be electronically equipped to communicate with one another and with the users, becoming an integral part of the Internet \cite{Zanella_Bui_Vangelista_Zorzi_2014}. 

Even though smart transportation refers to using smart technologies to improve citizens' mobility experience, its benefits certainly go beyond that. For instance, smart transportation also provides the infrastructure to enable other transport-related initiatives, such as predictive systems for collision avoidance, traffic monitoring, and real-time traveler information. In this context, a system for travel time estimation, such as the one herein proposed, may be regarded as a beneficiary and a contributor to the realization of smart transportation.

\begin{figure}[!t]
\centering
\includegraphics[width=\columnwidth]{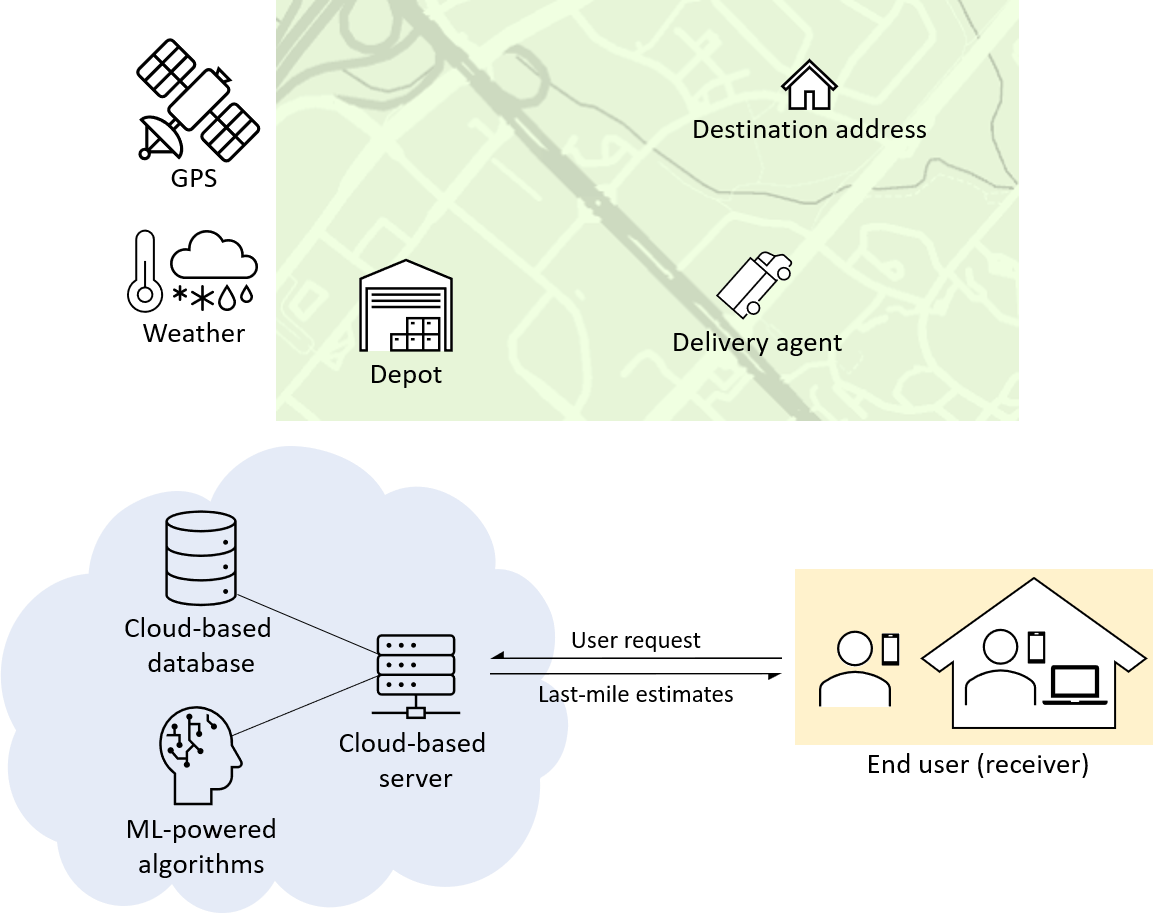}
\caption{The typical scenario for last-mile delivery is presented at the top, in which a parcel is taken from a local depot to a designated address by a delivery agent. At the bottom, a diagram illustrates the deployment of the system in an IoT scenario, and the relation between end-user and cloud infrastructure.}
\label{system_architecture}
\end{figure}

Figure \ref{system_architecture} illustrates the typical scenario for last-mile delivery, in which a parcel is scanned as \textit{out-for-delivery} in a local depot, collected by an agent, who then delivers it at the designated address. During the process, relevant data is recorded, such as depot geographical coordinates (GPS) and destination postal codes. Weather information is also available, including records on daily temperature, precipitation of rain and snow, and the amount of snow in the ground.

In the same figure, we illustrate the deployment of such a system in an IoT-enabled scenario. A diagram presents an end-user (receiver) perspective, who should be able to request the status of their parcel at any time or place, especially during the expected day of delivery. A cloud-based server, which works in conjunction with a database, would then process the user request, feed it into an appropriate machine learning inference algorithm, and return the last-mile estimates to the user. {\reviewcolor We, therefore, envision, as a potential immediate application of this solution, the development of a tracking app that would allow the end-user to anticipate the arrival of their parcel on the day of delivery. The solution could also be applied to other data-driven travel time scenarios, such as urban transportation services.}

{\reviewcolor Finally, given the incorporation of a learning module as part of this IoT system, it is worth noting that such contextualization could be expanded to that of Cognitive IoT (CIoT), a broader paradigm that encompasses not only the connection of physical objects as IoT devices but also the infusion of intelligence into the physical world through them \cite{Sassi_Jedidi_Fourati_2019}. In spite of that, we adopt the general IoT paradigm for purposes of contextualization, as described in the following section.}

\subsection{IoT Architecture}

\begin{figure}[!t]
\raggedleft
\includegraphics[width=.8\columnwidth]{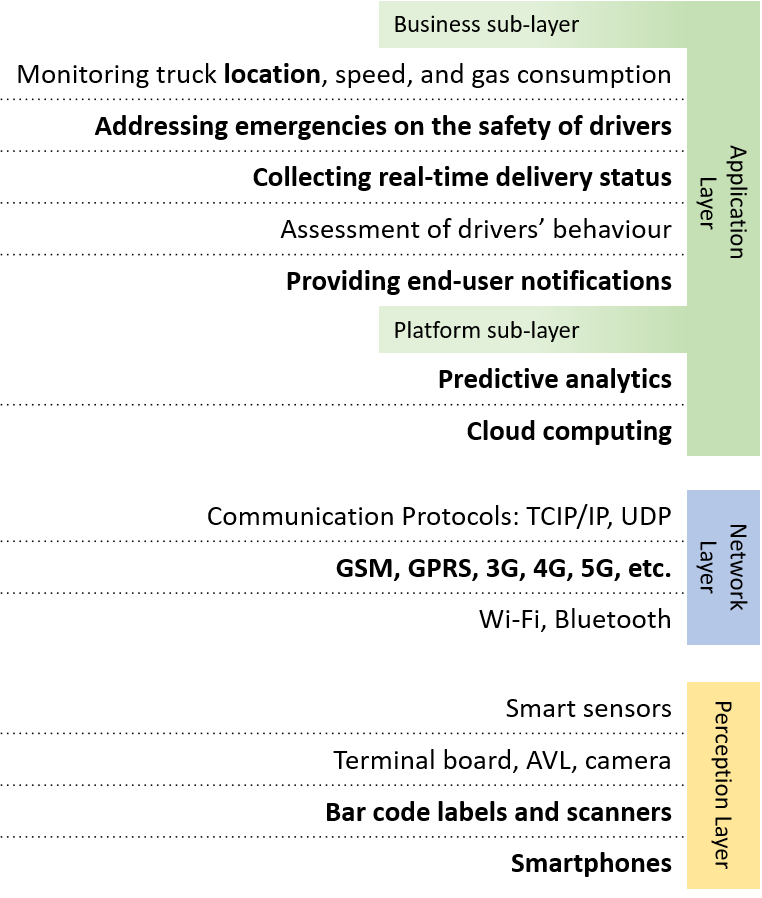}
\caption{An IoT 3-layered architecture is shown: the layer descriptions are adapted from \cite{luo2019new} to the context of the last-mile parcel delivery problem, highlighting some major points for each layer. We highlight in bold the terms relevant to the context of our last-mile delivery solution.}
\label{iot_architecture}
\end{figure}

Although a number of flexible layered architectures have been proposed for IoT standardization, a reference model has not yet been established \cite{al2015internet}. Still, in its most basic modeling, the IoT architecture generally comprises three layers of abstraction. This three-layer architecture consists of the Perception Layer, the Network Layer, and the Application Layer. In the following, we present them according to \cite{luo2019new} while placing them in the context of our problem. Finally, Figure \ref{iot_architecture} summarizes the IoT layered architecture.

\subsubsection{Perception Layer} 
This layer represents the information source of the IoT, that is, sensors and data collection devices. For example, the terminal board in delivery trucks enables the tracking of relevant data such as speed and gas consumption. Also, Automatic Vehicle Location (AVL) enables reliable transmission of current vehicle location. This layer could also include cameras or barcode labels used by delivery agents, besides any additional devices required for communication with the depot or the addressee, such as smartphones. Our solution will mainly exploit this layer by providing delivery agents with portable scanners associated with smartphones. With that, the agents would scan the parcels at the moment of leaving the depot, as well as at their final destinations, therefore generating the timestamps needed as inputs to our models. Each scan would also provide the respective GPS coordinates collected on their smartphones. Moreover, the agent would still be able to provide user notification in case of unforeseen events relative to the delivery.

\subsubsection{Network Layer} 
This layer essentially transmits information from the Perception to the Application Layer. It is established based on the current mobile telecommunication and internet infrastructure. In terms of wireless communication, examples include Global System for Mobile Communication (GSM), General Packet Radio Service (GPRS), and the 4th Generation communication (4G) or 5th Generation (5G), which can allow for long-distance transmission of vehicular data. Besides, private wireless communication such as WiFi and Bluetooth can be used for communicating sensors within the trucks, for example. For our proposed last-mile delivery solution, this layer would simply make use of the mobile network in the delivery agents' smartphones in order to transmit the required coordinates and timestamps back to their respective depots.

\subsubsection{Application Layer} 
This layer receives information from the Perception Layer and processes it into useful applications for staff and customers. It can be subdivided into platform and business sub-layers. The platform sub-layer comprises all algorithms used, for example, for data mining and inference, with cloud computing support. The business sub-layer holds the applications per se. Regarding the platform sub-layer for our last-mile delivery solution, our deep learning models for travel time estimation will operate as the predictive analytics engine. Our models and data would be stored on the cloud and used for inference by customer demand as well as for training the models. On the other hand, the business sub-layer comprises not only the user device but also the driver's smartphone. Based on its definition, we place the driver smartphone in both Perception in Application layers. The driver device would automatically report truck location, speed, and driving behavior, while also notifying emergencies regarding the safety of drivers. The user device in turn would allow for collecting real-time delivery status and notifications to customers, who would be able to query updates at any time using any device connected to the Internet. 

\section{Method and Experiments}

\subsection{Problem Formulation}

We tackle the problem of last-mile parcel delivery time estimation as an \textit{Origin-Destination Travel Time Estimation} (OD-TTE) problem, that is, we aim to predict travel time based on start and end locations, not relying on the actual route taken. {\reviewcolor Given dataset $\mathcal{D}$ containing $N$ samples $d_i$ defined as:
\begin{eqnarray}
    \label{eq_dataset_1}
    \mathcal{D} &=& \big\{d_i\ |\ i \in [1,N] \big\} \\
    \label{eq_dataset_2}
    d_i &=& (o_{x_i}, o_{y_i}, o_{t_i}, d_{x_i}, d_{y_i}, w_i, t_i),
\end{eqnarray}
where the \textbf{last-mile delivery record} tuple $d_i$ comprises:
\begin{itemize}
    \item the longitude/latitude coordinates for the depot ($o_{x_i}, o_{y_i}$);
    \item the coordinates for the postal code destination ($d_{x_i}, d_{y_i}$);
    \item the timestamp $o_{t_i}$ relative to the out-for-delivery scan;
    \item a vector $w_i$ encoding weather conditions;
    \item the registered delivery time \textit{target} $t_i$, the time elapsed from $o_{t_i}$ up to the moment when the last scan happened.
\end{itemize}
}
\newtheoremstyle{definition}
{}{}{\itshape}{}{\bfseries}{.}{ }
{\thmname{#1}\thmnote{ (#3)}}
\theoremstyle{definition}
\newtheorem{defn}{Problem (OD Travel Time Estimation)}[]
\begin{defn}
Assuming the availability of a dataset of past deliveries $\mathcal{D}$ (according to Equation \ref{eq_dataset_1}), the end goal of OD-TTE is to leverage $\mathcal{D}$ so that, given a parcel query $q = (o_{x}, o_{y}, o_{t}, d_{x}, d_{y}, w)$, its correspondent travel (delivery) time $t$ can be predicted with minimal error.
\end{defn}

{\reviewcolor Our method consisted of studying the effectiveness of different architectural categories of CNNs for tackling this problem, which are described in the next sections. We aim to learn a model as a parametric map defined by a deep neural network. Even though we mathematically describe the problem, architectures, and experiments involved in this work, a theoretical study is outside the scope of this work.}

\subsection{VGG-based Architectures} 
\begin{figure}[!t]
\centering
\includegraphics[width=.8\columnwidth]{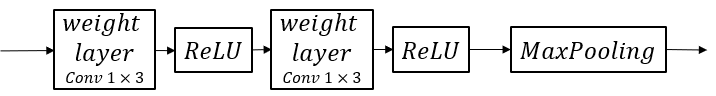}
\caption{VGG building block ($1D$) \cite{simonyan2014deep}.}
\label{vgg_block}
\end{figure}

The first class of convolutional networks explored is based on VGG modules \cite{simonyan2014deep}. The VGG block generally comprises a number of convolutional layers followed by a pooling layer. Figure \ref{vgg_block} shows a $1D$ version of the VGG block, formed by two convolutional ReLU-activated layers with a kernel of $3$, and a max-pooling layer with a stride of $2$.

\begin{figure}[!t]
\centering
\includegraphics[width=1\columnwidth]{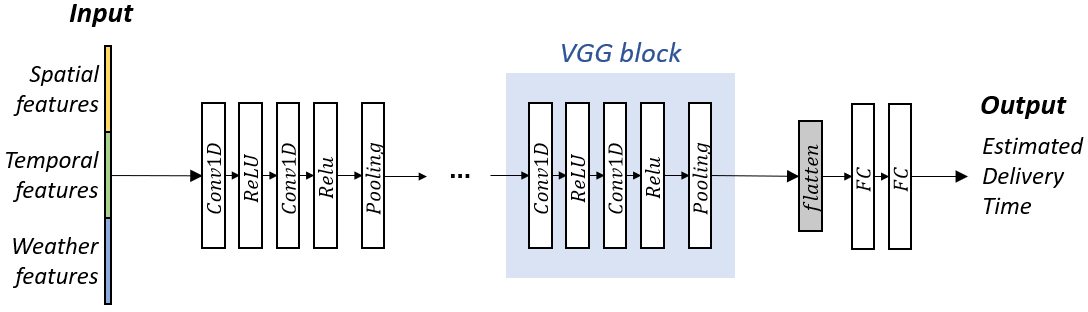}
\caption{A general diagram is presented describing the architecture of the VGG-based models explored, according to which we experiment variations on the number of VGG blocks with increasing depth.}
\label{vgg_arch}
\end{figure}

By combining such blocks, we build a series of VGG-based neural networks of varying depths. Figure \ref{vgg_arch} presents a general diagram of the VGG architecture utilized. On the input end, we concatenate the encoded features, which feeds a series of VGG blocks of increasing depth, allowing the model to learn complex representations in the data. Then, this deep representation is flattened into a one-dimensional tensor and fed to a fully-connected two-layered network, which outputs the estimated prediction of delivery time. 

The first and shallowest reported VGG-based model has $3$ blocks of increasing depth, a total of $6$ convolutional layers and $3$ pooling stages. In principle, the number of pooling layers, and therefore VGG blocks, is limited by the dimensionality of the input. In our case, for an input dimension of $12$, that limit would be $3$ (for a stride of $2$). To account for that when building deeper VGG models, we distribute the pooling layers evenly, while prioritizing earlier layers if needed. We further expand the experiments to deeper models containing up to $10$ VGG blocks, therefore with $6$ to $20$ convolutional layers.

\subsection{ResNet-based Architectures} 

\begin{figure}[!t]
\centering
\includegraphics[width=.7\columnwidth]{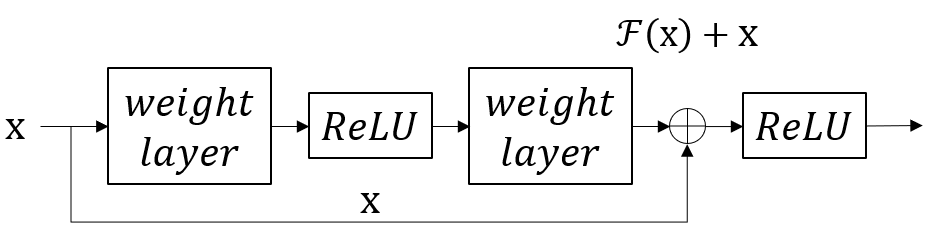}
\caption{Residual learning unit, as introduced in \cite{He_2016_CVPR}. We follow the original approach of aggregating the skip connection before the last ReLU activation function.}
\label{residual_block}
\end{figure}

Residual learning explores the building of computational units that learn residual functions of the input, in contrast to learning unreferenced mappings \cite{He_2016_CVPR}. This helps to mitigate the problem of vanishing gradients that affects deeper networks. Practically, a shortcut (or skip) connection is associated with a two-layer feed-forward ReLU-activated neural network, resulting in the ResNet block, the residual learning unit used in ResNet architectures \cite{He_2016_CVPR}, as shown in Figure \ref{residual_block}.

\begin{figure}[!t]
\centering
\includegraphics[width=1\columnwidth]{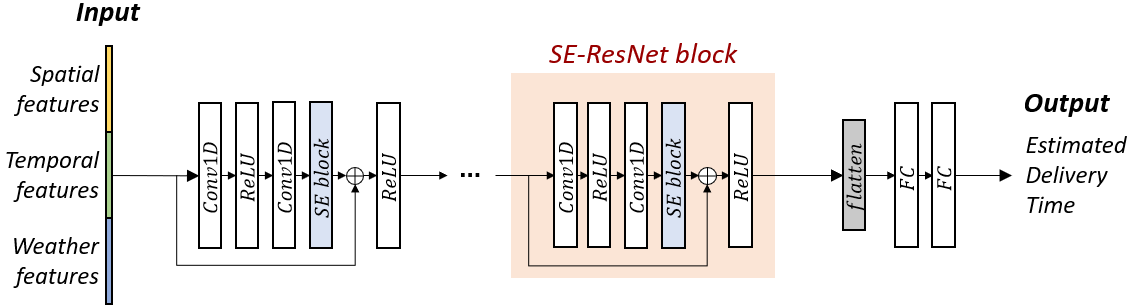}
\caption{A general diagram is presented describing the architecture of our deep residual models. Likewise, the number of blocks is not specified since we experiment with different instances of varying depth.}
\label{resnet_arch}
\end{figure}

The second class of convolutional neural networks investigated is based on this concept, i.e. residual modules (ResNet). Their innovation relates to the presence of skip connections serving as auxiliary paths for gradient flow. It is worth noting that ResNet blocks by definition do not presume pooling. Figure \ref{resnet_arch} shows the counterpart architectural diagram for the ResNet models built. Similarly, we experiment with models of varying depth, ranging from $3$ to $10$ residual blocks, that is, models with $6$ to $20$ convolutional layers.

\subsection{Squeeze-and-Excitation Architectures}

The Squeeze-and-Excitation (SE) block \cite{Hu_2018_CVPR} is an approach for modeling interdependencies between channels in convolutional networks. While prior architectures usually learn spatial and channel-wise feature maps jointly, SE blocks attempt to learn them independently. It achieves that by collapsing spatial dimensions, through pooling, to model channel-specific properties and re-calibrate channel-wise responses accordingly.

One advantage of the SE block is its ease of being integrated into neural network modules with no significant additional computational cost. As a result, they have been used in a variety of applications. This block has benefited CNN models on problems using 2D image data \cite{8882973, 8803000, 8808853, 8447284}, 3D volumetric data \cite{Rickmann_2020, Han_Wei_Zhou_Hong_Zhang_Yang_2020, 9043526, 8998244}, and on solutions on 1D signals, such as music, speech \cite{lee2017raw, SampleCNN_2019, unsup_ASR_SE_2019, han2020contextnet}, and EEG \cite{se_eeg_2020}.

\begin{figure}[!t]
\centering
\includegraphics[width=\columnwidth]{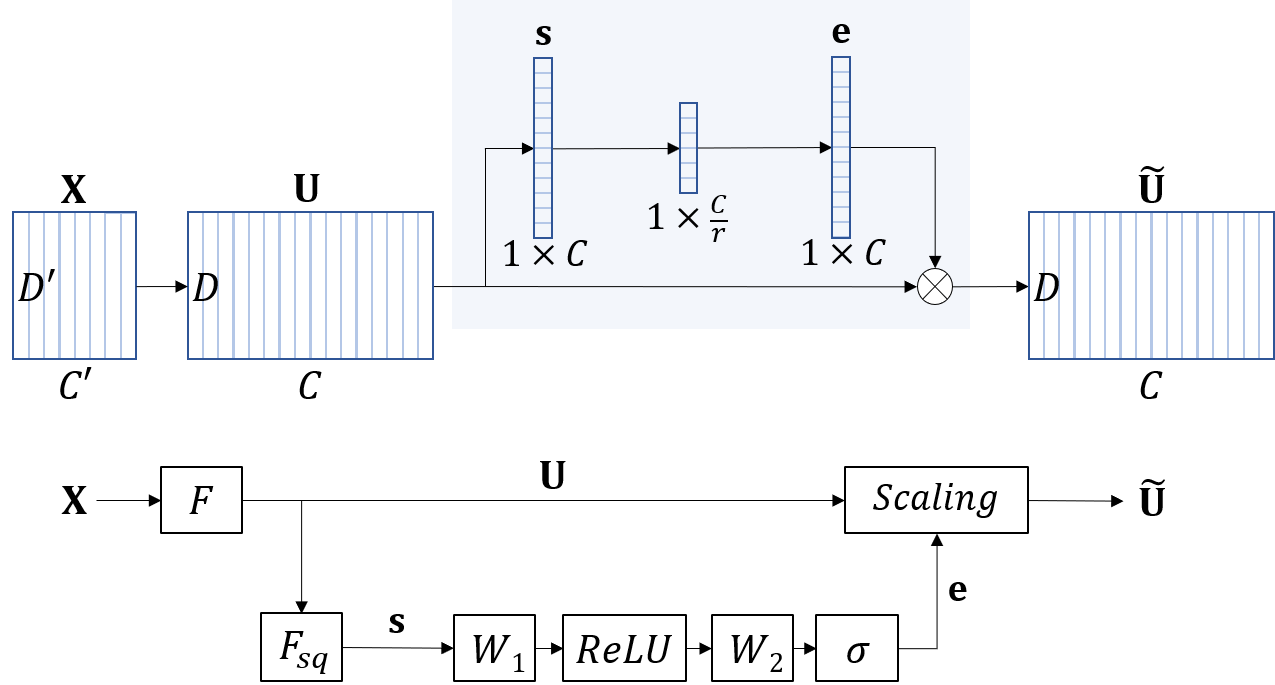}
\caption{A diagram on the squeeze-and-excitation block to 1D data is shown.}
\label{se_block}
\end{figure}

The SE block can be used to re-calibrate the feature map from any transformation $F : \mathbf{X} \to \mathbf{U}$, $\mathbf{X} \in \mathbb{R}^{D'\times C'}$, $\mathbf{U} \in \mathbb{R}^{D\times C}$, so $F$ could represent a regular convolution, a residual module, an inception module, etc. In the following, the main components of an SE block adapted to 1D inputs for our problem space are presented. Figure \ref{se_block} summarizes the SE block pipeline: a diagram at the top illustrates how the input tensor dimensions change through each stage of the block; at the bottom, a complementary schema highlights the transformations that take place.

\textbf{Squeeze operator.} The first operator $F_{sq}$ in the SE block squeezes the input along its spatial dimension, collapsing the tensor so that only the channel dimension remains. Therefore, the squeeze operator outputs a channel-wise vector $\mathbf{s} \in \mathbb{R}^{C}$. Generally, global average pooling is used, so each channel component $\mathbf{u}_c$ is mapped to a scalar $s_c$ according to:
\begin{equation}
    s_c = F_{sq}(\mathbf{u}_c) =\frac{1}{D} \sum_{i=1}^D \mathbf{u}_c(i)\ .
\end{equation}

\textbf{Excitation operator.} The second operator $F_{ex}$ in the SE block is intended to exploit the ``squeezed'' representation $\mathbf{s}$ in order to capture the importance of each channel regardless of the spatial distribution. That is achieved by a two-layer fully connected (FC) neural network given by:
\begin{equation}
    \mathbf{e} = F_{ex}(\mathbf{s}) = \sigma\left( \mathbf{W}_2 \mathrm{ReLU}( \mathbf{W}_1 \mathbf{s})  \right) ,
\end{equation}
where $\mathbf{e} \in \mathbb{R}^{C}$, $\mathbf{W}_1 \in \mathbb{R}^{C/r \times C}$, $\mathbf{W}_2 \in \mathbb{R}^{C \times C/r}$. The first layer reduces the depth of $\mathbf{s}$ by a factor of $r$, while the second restores it back to $C$. A sigmoid activation function at the last layer assures that channels can be independently excited.

\textbf{Feature scaling.} In the final step, each channel of $\mathbf{U}$ is re-scaled by the correspondent factor in the vector $\mathbf{e}$, an independent re-calibration of features that allows the model to select most relevant ones. The following equation denotes a channel-wise multiplication:
\begin{equation}
\mathbf{\widetilde{U}} = \mathbf{e} \cdot \mathbf{U}
\end{equation}

\begin{figure}[!t]
\centering
\includegraphics[width=.8\columnwidth]{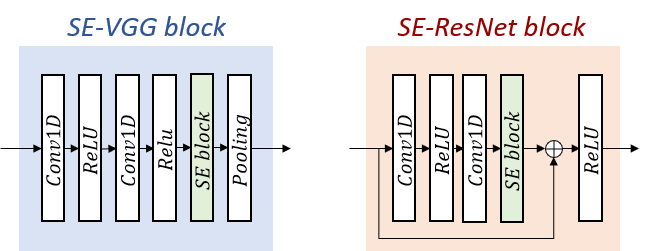}
\caption{Diagrams for the squeeze-and-excitation versions of the VGG and ResNet blocks. The insertion point for SE-VGG and SE-ResNet are respectively before pooling and before aggregating the skip connection.}
\label{se_blocks}
\end{figure}

In this study, we incorporate the SE block into our best VGG and ResNet models as an attempt to better capture more intricate relationships between the learned feature maps. For the VGG models, we insert the SE block after the non-linearity, right before the pooling operator, while in the ResNet architectures, the SE block is best placed before the aggregation with the skip connection. We illustrate the augmented versions of each block in Figure \ref{se_blocks}.

\subsection{Training and Implementation Details}

For training our model we utilized the Adam optimizer \cite{adam}, and the \textit{Mean Squared Error} (MSE) as loss function. Considering a given set of $N$ samples $y_i$ of delivery times and $N$ associated predictions $f_i$, the MSE is defined by:
\begin{eqnarray}
    \label{equation_mse}
    MSE = \frac{1}{N} \sum_{i=1}^N \left| y_i - f_i  \right|^2.
\end{eqnarray}

We monitor the loss over the validation set and apply early stopping if no improvement happens after $25$ consecutive epochs. We set an initial learning rate of $10^{-4}$ and reduce it by half every $40$ epochs. We utilize a batch size of $64$ samples. We implement our models using the Keras API with TensorFlow \cite{abadi2016tensorflow} backend, using an Nvidia RTX 2080 Ti GPU.

\subsection{Dataset}
We utilize a real-world dataset on last-mile parcel delivery provided by Canada Post, the main postal operator in Canada. The dataset holds $3,253,252$ deliveries within the Greater Toronto Area (GTA), from January to June of 2017. There are $72$ different depot locations in the data, covering $83,730$ delivery points. Geographical coordinates (longitudes and latitudes) are given for the depots, but, due to privacy, delivery addresses are limited to postal codes, from which coordinates are obtained. Timestamps are also available, indicating the date and time when the \textit{out-for-delivery} scan and the \textit{delivered} scan happened, from which delivery time is computed.

\begin{figure}[!t]
\centering
\includegraphics[width=0.8\columnwidth]{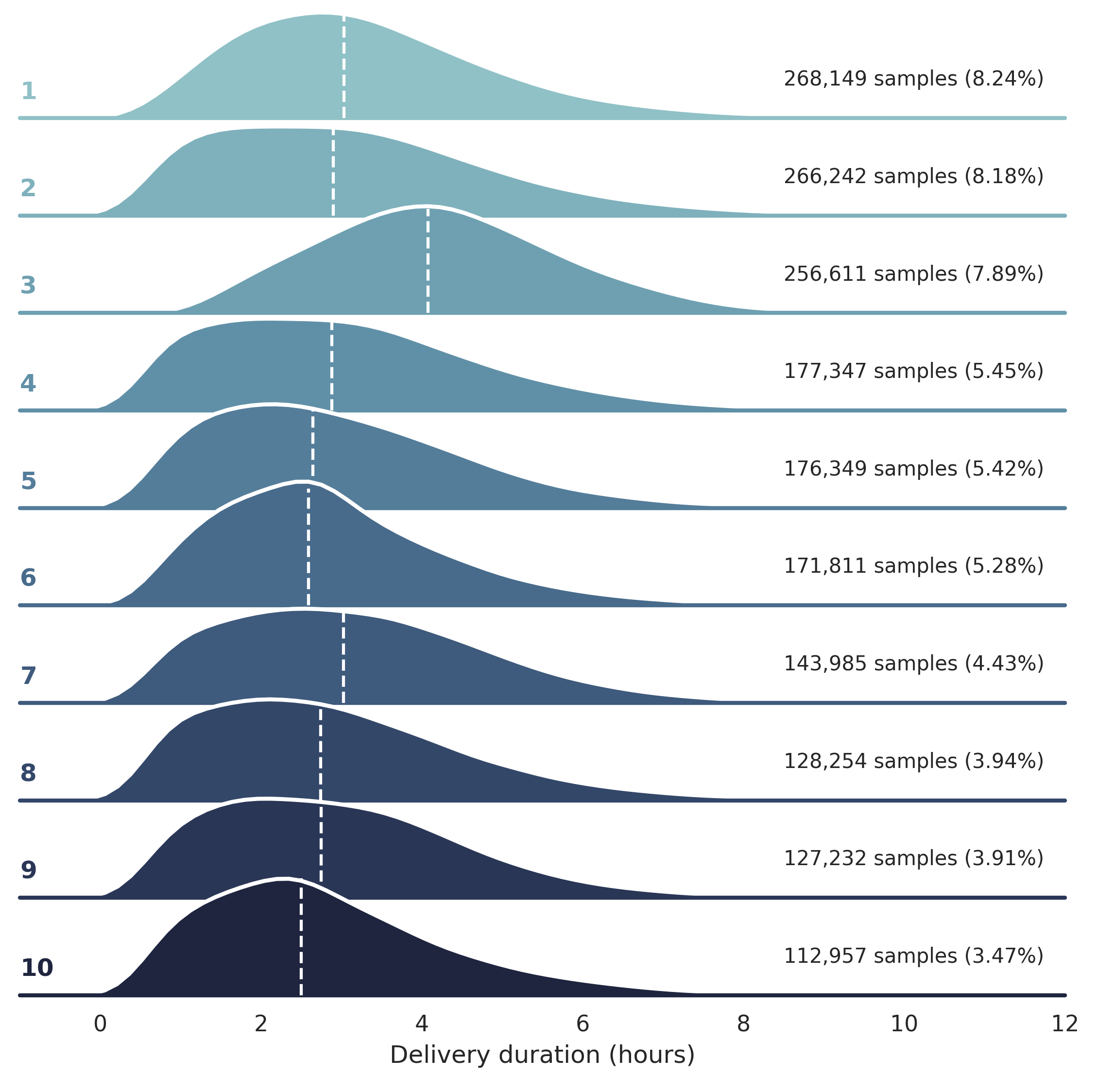}
\caption{The distribution of delivery durations for the 10 busiest depots, which jointly cover $56.22\%$ of the data. The busiest location is at the top in the lightest color. For reference, dashed white segments show their medians.}
\label{target_across_depots}
\end{figure}

The last-mile parcel delivery problem shows a clear imbalance between the number of origins and destinations, unlike other OD-based problems such as predicting taxi trip durations, for example. Since there are considerably fewer depots than delivery addresses, each depot usually covering a defined region, it is reasonable to expect spatio-temporal patterns to vary across different depots as they relate to different traffic and road conditions. Figure \ref{target_across_depots} illustrates the variations on the distribution of delivery time for each of the $10$ busiest depots.

\begin{figure}[!t]
\centering
\includegraphics[width=\columnwidth]{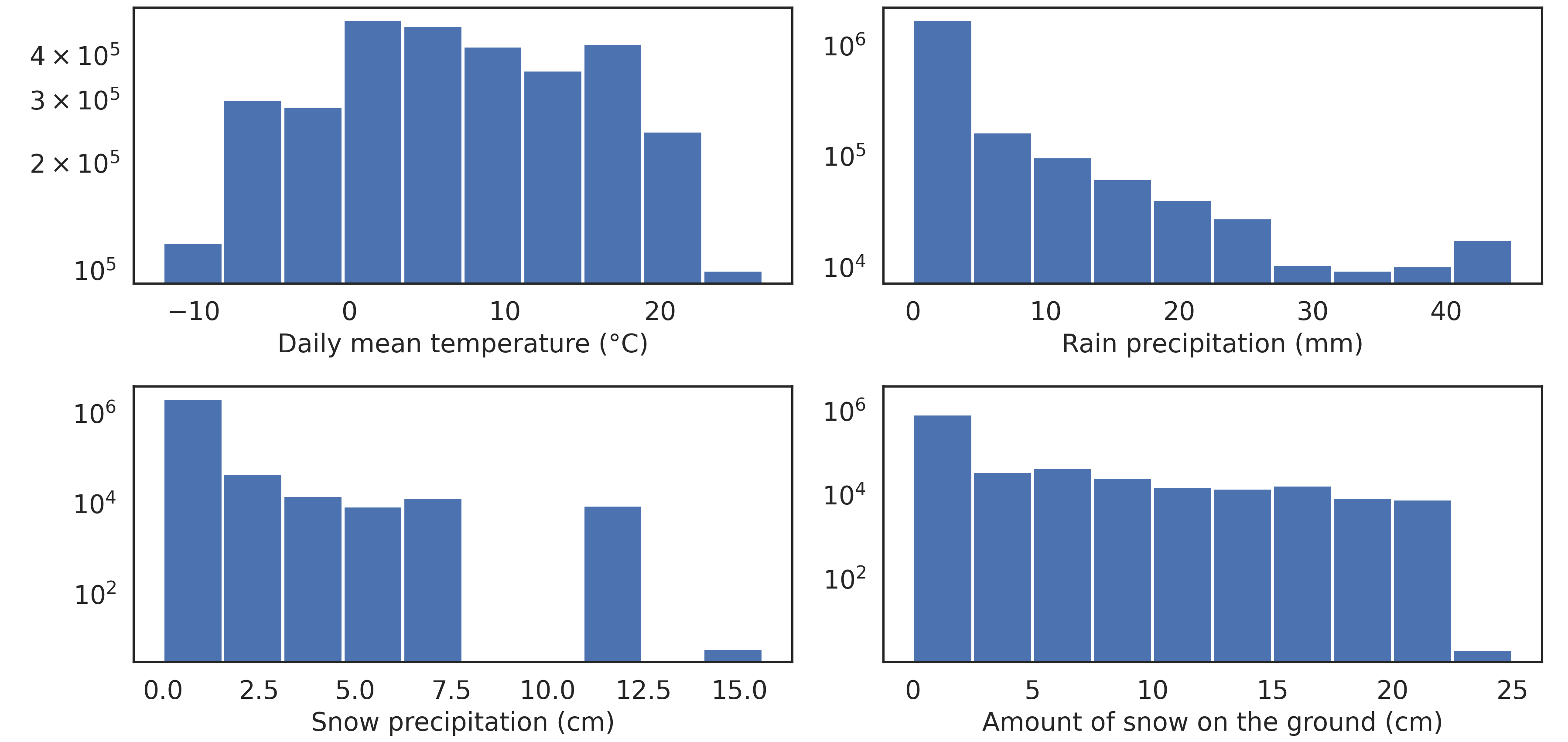}
\caption{Distribution of weather features is shown. The histograms are plotted in logarithmic scale.}
\label{weather_distribution}
\end{figure}

Weather information is also used, as it configures a critically relevant factor for delivery times. Besides, the period spanned covers diametrically opposite climate conditions (e.g. January vs. June). The features used are daily temperature, rain and snow precipitation, and the amount of snow on the ground. Figure \ref{weather_distribution} shows how these features are distributed in the data. The plot on the top-left shows the daily temperature, which ranges from $-10^{\circ}$ to $30^{\circ}$. This figure shows wide distributions of different weather conditions, indicating that deliveries have occurred under many different situations, contributing to the hardness of this dataset. We observe that rain precipitation has been as high as $40$ millimeters, and snow precipitation has reached up to $16$ centimeters. While local measurements might have been different, the reported values are daily averages, computed across multiple weather stations. Finally, Table \ref{table_data_summary} summarizes some high-level information on the GTA dataset.

\begin{table}[!t]
\caption{GTA Dataset Summary}
\label{table_data_summary}
\centering
\begin{tabular}{ll}
\hline
Statistic & Value \\ \hline\hline
Number of samples & $3.25$M \\
Number of depots & $72$ \\
Number of delivery points & $83$k \\
Average delivery time & $3.19$ h \\
Median delivery time & $2.96$ h \\
Variance of delivery time & $2.88$ \\
\hline
\end{tabular}
\end{table}

\subsection{Input Features}
In terms of coordinates, especially for final destinations, we quantize the latitudes and longitudes to a coarser resolution to alleviate GPS noise and to cluster neighboring locations together. Next, we normalize them based on a geographical bounding box enclosing the GTA. The Haversine distance, which measures the distance between two points on Earth based on their coordinates, is computed for each delivery and further normalized as well. From the out-for-delivery timestamp, we extract hour-of-day, day-of-week, and the week itself. The daily temperature, precipitation of both rain and snow, and the amount of snow on the ground are also used. All temporal and weather features are normalized within the range of $0$ to $1$. Finally, we concatenate all aforementioned features into a $12$-dimensional input vector.

\subsection{Testing Protocol}
We randomly split the data to a ratio of $70\%$ for training and $30\%$ for testing. In order to measure performance, we use metrics that are common to travel time estimation problems \cite{wang_simple,Bertsimas_2019,2017arXiv171004350J,li2018multi}. Besides the loss function (MSE), we report its root (RMSE) as well as the Mean Absolute Error (MAE), the Mean Absolute Percentage Error (MAPE), and the Mean Absolute Relative Error (MARE). Given a set $\{y_1, \ldots, y_N\}$ of $N$ delivery time samples and a correspondent set of predictions $f_i$, the referred metrics are defined in the following.
\begin{eqnarray}
    MAE  &=& \frac{1}{N} \sum_{i=1}^N \left| y_i - f_i  \right|\\
    \label{metrics_mape}
    MAPE &=& \frac{1}{N} \sum_{i=1}^N \left| \frac{y_i - f_i}{y_i}   \right|\\                  
    \label{metrics_mare}
    MARE &=& \frac{ \sum_{i=1}^N \left| y_i - f_i \right|}{\sum_{i=1}^N \left| y_i \right|} 
\end{eqnarray}

Additionally, we measure the window of error ($\pm EW$) that includes a certain percentage of the delivery predictions. For a desired percentage of $p$, $EW$ can be found by:
\begin{eqnarray}
\frac{1}{N}\sum_{i=1}^N u(EW - \left| y_i - f_i \right|) = p\ ,
\end{eqnarray}
where $u(\cdot)$, the step function, counts the number of samples whose absolute errors are within $\pm EW$ hours. In particular, we report $EW_{90\%}$. This metric was indicated by Canada Post as a common evaluation metric in that organization.

\subsection{Benchmarks}

\subsubsection{Random Forest}
we explored random forests as a representative of classical ensemble methods. We define a grid search over their main hyperparameters, constraining exaggerated growth and watching for overfitting.
 
\subsubsection{Multi-Layer Perceptrons}
we also explored neural networks based on Multi-Layer Perceptrons (MLPs). We report two instances of $2$ and $5$ ReLU-activated layers, each with $50$ units, trained with early stopping based on validation loss.

\subsubsection{SB-TTE \cite{wang_simple}} 
this efficient OD solution works by grouping similar trips based on the distances between their origins and their destinations. A scaling factor based on \textit{actual} travel distances accounts for different traffic conditions.

\subsubsection{ST-NN \cite{2017arXiv171004350J}} 
a cascade of two MLP modules jointly trained for the sum of their losses. The first estimates OD traveled distance from GPS coordinates; the second combines temporal features with those learned in the first module to predict travel time. A particular temporal encoding is used.

\subsubsection{DNN \cite{deep_net_travel_2019}} 
also an MLP-based OD solution, this work predicts vehicle travel times using a trigonometric encoding for temporal features and an auxiliary model trained to estimate the actual traveled distance over a grid map of the city. 

\newtheoremstyle{definition}
{}{}{}{}{\bfseries}{.}{ }
{\thmname{#1}\thmnote{ (#3)}}
\theoremstyle{definition}

\newtheorem{pargraph}{A remark on referenced baselines}[]
\begin{pargraph}
Adaptations were necessary when re-implementing the solutions found in the literature, mostly because the closest approaches, although OD-based, would still indirectly rely on the availability of the actual distance traveled by the driver \cite{wang_simple,2017arXiv171004350J, deep_net_travel_2019}, information usually available in public datasets.
\end{pargraph}

\section{Results and Analysis}

\subsection{Evaluating VGG Architectures}

\begin{figure}[!t]
\centering
\includegraphics[width=0.7\columnwidth]{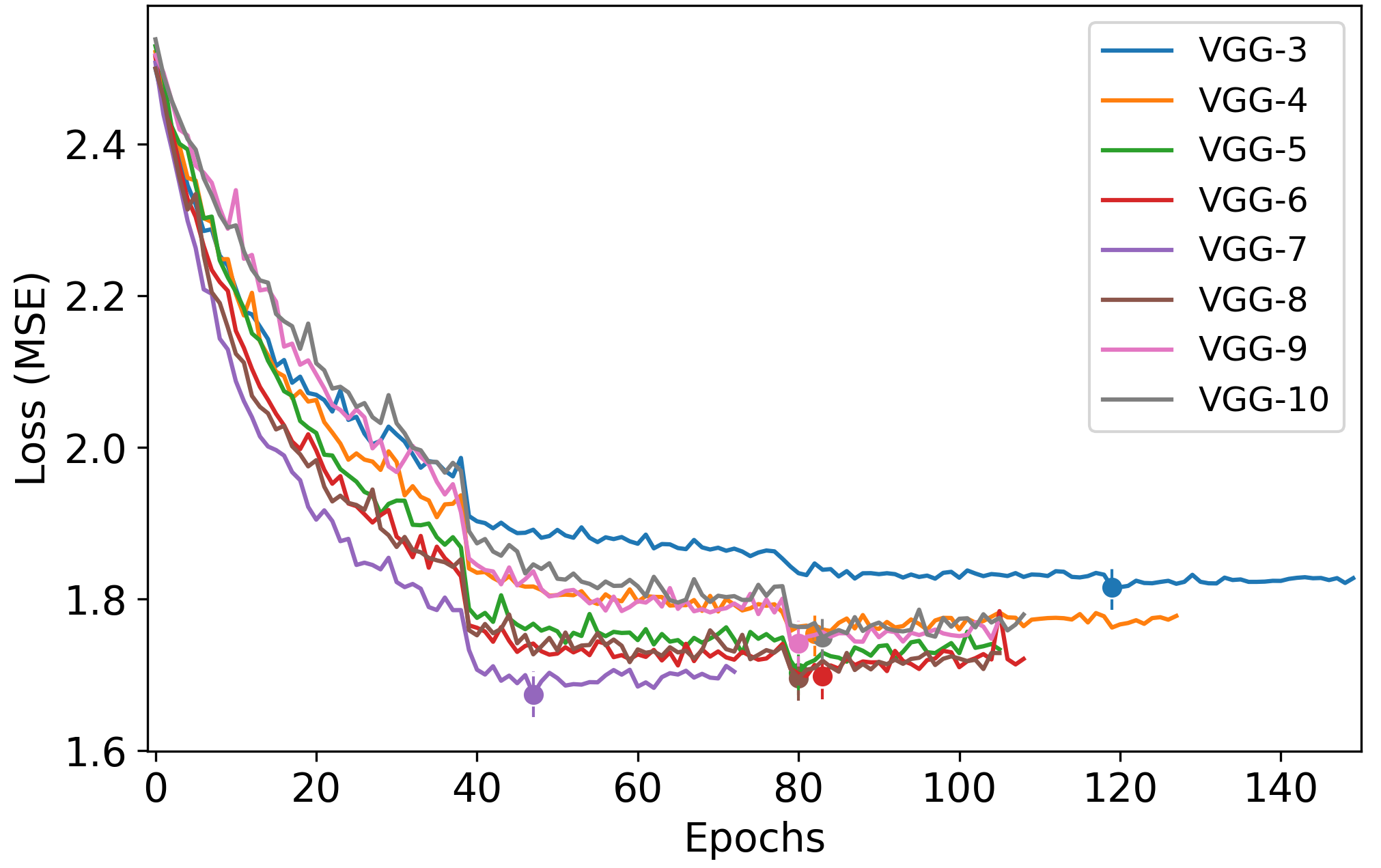}
\caption{Comparing the validation loss (MSE) for all VGG architectures. Small circles mark the epochs after which they stop improving.}
\label{vgg_losses}
\end{figure}

We now present the performance results on the VGG architectures over the GTA dataset. We varied the number of VGG blocks to identify the best trade-off between performance and complexity. Table \ref{vgg_table} enlists all metrics computed over the validation set for the VGG models with the depth varying from $3$ up to $10$ blocks. Regarding performance, even though we observe an improvement as we explore additional blocks, it deteriorates on the deeper instances, particularly after VGG-7 ($14$ layers). This could be due to overfitting or the vanishing gradient problem. We display the validation losses during training in Figure \ref{vgg_losses}, in which a similar trend can be observed. Accordingly, VGG-6 displays the best performance: although it does not have the lowest loss, it shows the best overall results when considering all metrics.

\begin{table}[!h]
\caption{Evaluating the effect of depth in VGG architectures}
\label{vgg_table}
\centering
\begin{adjustbox}{width=0.8\columnwidth}
\begin{tabular}{l|l|l|l|l|l|l}
\hline
& MSE & RMSE & MAE & $\text{EW}_{90\%}$ & MAPE & MARE \\
\hline
\hline
VGG-3     & 1.8273 & 1.3518 & 0.9789 & 2.0117 & 44.04 & 30.71       \TBstrut  \\
VGG-4     & 1.7542 & 1.3245 & 0.9417 & 1.9527 & 41.83 & 29.55       \TBstrut  \\
VGG-5     & 1.7028 & 1.3049 & 0.8961 & 1.8749 & 39.45 & 28.12       \TBstrut  \\
VGG-6     & 1.6979 & 1.3030 & 0.8867 & 1.9160 & 39.36 & 27.82       \TBstrut  \\
VGG-7     & 1.6743 & 1.2939 & 0.9179 & 1.8864 & 41.46 & 28.80       \TBstrut  \\
VGG-8     & 1.6955 & 1.3021 & 0.9003 & 1.8623 & 40.35 & 28.25       \TBstrut  \\
VGG-9     & 1.7411 & 1.3195 & 0.9363 & 1.8931 & 42.17 & 29.38       \TBstrut  \\
VGG-10    & 1.7494 & 1.3227 & 0.9408 & 1.9258 & 41.94 & 29.52       \TBstrut  \\ \hline
\end{tabular}
\end{adjustbox}
\end{table}

\subsection{Evaluating ResNet Architectures}

\begin{figure}[!t]
\centering
\includegraphics[width=0.75\columnwidth]{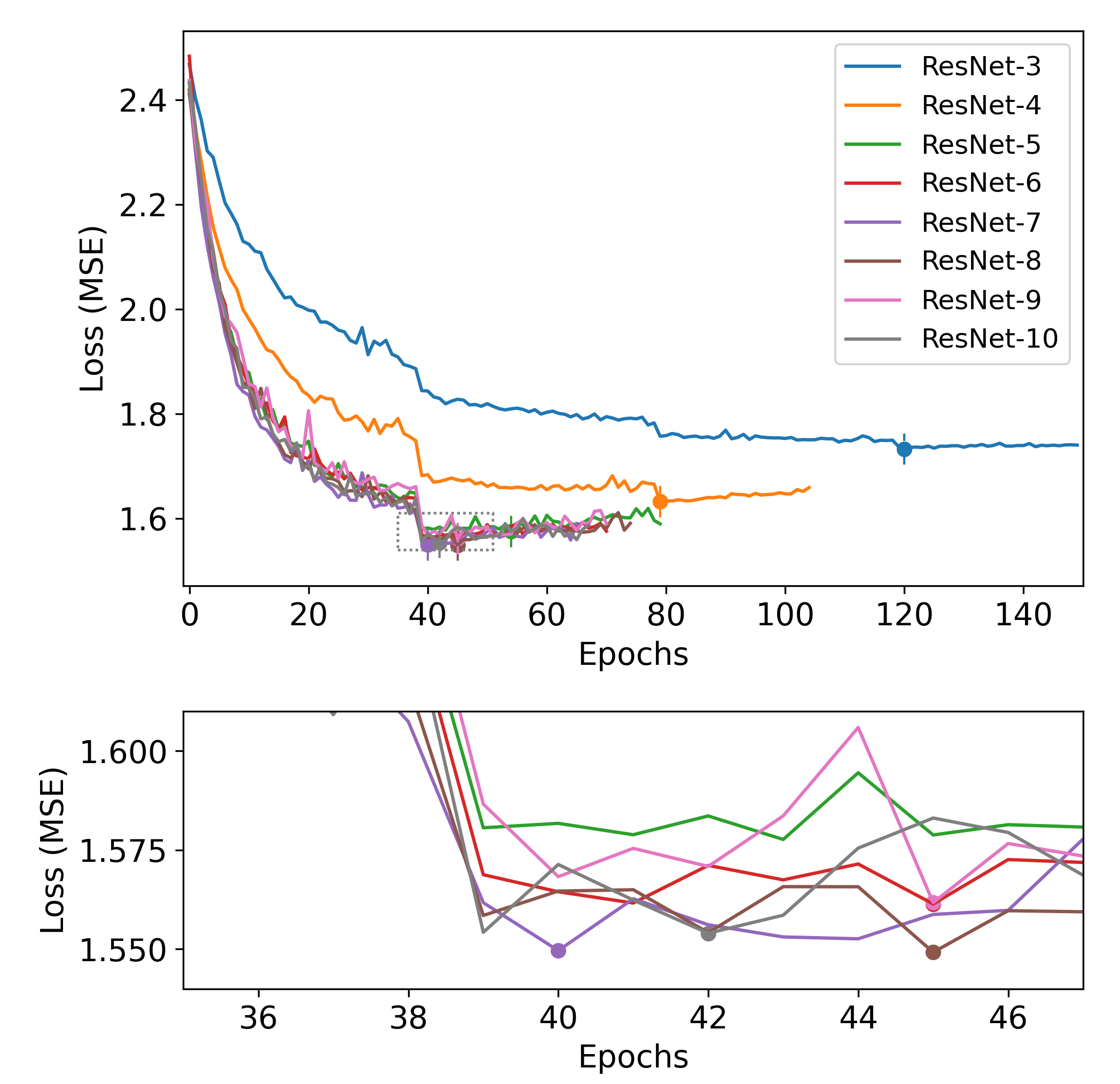}
\caption{Comparing the validation loss (MSE) for all ResNet architectures. In the bottom, we zoom in on the region where losses reach their minimum. Small circles mark the epochs after which they stop improving.}
\label{resnet_losses}
\end{figure}

Next, we evaluate the different number of layers in our ResNet architectures. Table \ref{resnet_table} shows the results for instances with depth varying from $3$ up to $10$ ResNet blocks. Even though we can observe a generally monotonic decrease across all metrics as the models go deeper, the improvement diminishes. A similar trend is observed in the validation loss in Figure \ref{resnet_losses}. Small circles denote the epochs at which the models have reached their best performance, illustrated in the zoomed-in plot. Even though ResNet-8 reaches the lowest MSE, the best MAPE value, for example, was shown by ResNet-10. Metrics penalize errors differently and it is normal for them not to be always consistent, reason why we report six of them. 

The ResNets show a quite different evolution in comparison to the VGGs: the skip connections significantly help the learning process by providing direct paths for gradient flow during training. Accordingly, we infer that the decline in the performance of deeper VGGs may have been gradient-related.

\begin{table}[!h]
\caption{Evaluating the effect of depth in ResNet architectures}
\label{resnet_table}
\centering
\begin{adjustbox}{width=0.8\columnwidth}
\begin{tabular}{l|l|l|l|l|l|l}
\hline
& MSE & RMSE & MAE & $\text{EW}_{90\%}$ & MAPE & MARE \\
\hline
\hline
ResNet-3     & 1.7401 & 1.3191 & 0.9503 & 1.9569 & 42.88 & 29.81       \TBstrut  \\
ResNet-4     & 1.6320 & 1.2775 & 0.8911 & 1.8524 & 39.37 & 27.96       \TBstrut  \\
ResNet-5     & 1.5752 & 1.2551 & 0.8510 & 1.8419 & 37.28 & 26.70       \TBstrut  \\
ResNet-6     & 1.5614 & 1.2495 & 0.8516 & 1.8350 & 37.23 & 26.72       \TBstrut  \\
ResNet-7     & 1.5496 & 1.2448 & 0.8461 & 1.7942 & 36.61 & 26.55       \TBstrut  \\
ResNet-8     & 1.5492 & 1.2447 & 0.8404 & 1.7680 & 36.55 & 26.37       \TBstrut  \\
ResNet-9     & 1.5617 & 1.2497 & 0.8475 & 1.8248 & 36.95 & 26.59       \TBstrut  \\
ResNet-10    & 1.5540 & 1.2466 & 0.8411 & 1.7938 & 36.21 & 26.39       \TBstrut  \\ \hline
\end{tabular}
\end{adjustbox}
\end{table}

\subsection{Assessing the Squeeze-and-Excitation Augmentation}

We now report the results on the SE-based models. As discussed, the SE block is an architectural unit designed to \textit{enhance} the representational power of a model through feature re-calibration \cite{Hu_2018_CVPR}. Accordingly, we augmented VGG-6 and ResNet-8, our best VGG and ResNet instances, with SE blocks to assess their effect on the performance.

\begin{figure}[!t]
\centering
\includegraphics[width=0.75\columnwidth]{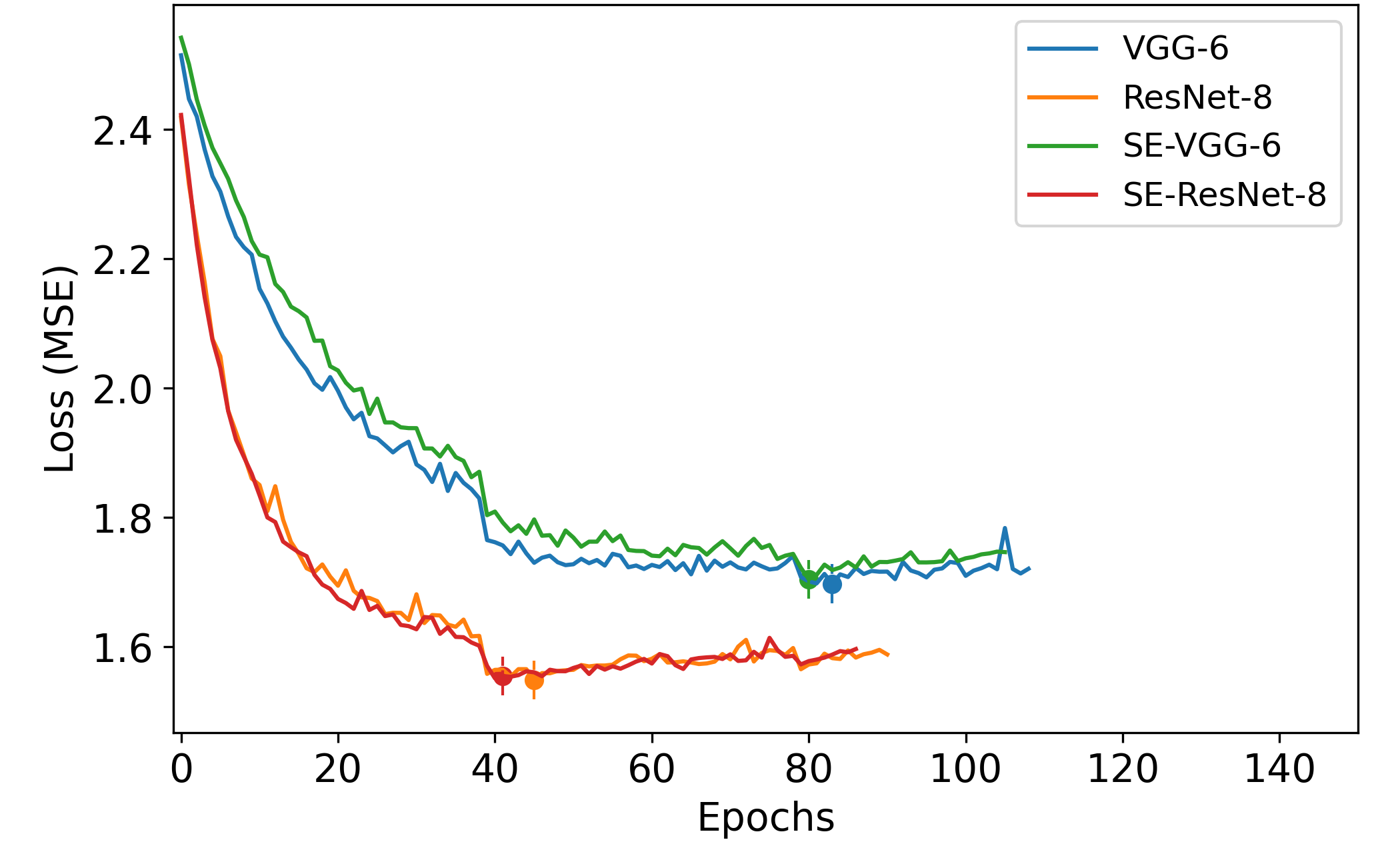}
\caption{Comparing the validation loss (MSE) for the best VGG and ResNet architectures (VGG-6 and ResNet-8) against their SE-augmented versions. Small circles mark the epochs after which they stop improving.}
\label{ses_losses}
\end{figure}

Table \ref{se_effect_table} shows the metrics for all four models. We observe that the insertion of SE blocks does not seem to improve performance. In fact, the evolution of losses in Figure \ref{ses_losses} for each pair is quite similar, especially for ResNet, and the SE blocks could actually be slightly worsening the performance.

\begin{table}[!h]
\caption{Evaluating the effect of squeeze-and-excitation augmentation}
\label{se_effect_table}
\centering
\begin{adjustbox}{width=0.8\columnwidth}
\begin{tabular}{l|l|l|l|l|l|l}
\hline
& MSE & RMSE & MAE & $\text{EW}_{90\%}$ & MAPE & MARE \\
\hline
\hline
VGG-6        & 1.6979 & 1.3030 & 0.8867 & 1.9160 & 39.36 & 27.82       \TBstrut  \\
SE-VGG-6     & 1.7048 & 1.3057 & 0.9003 & 1.8814 & 39.86 & 28.25       \TBstrut  \\ \hline
ResNet-8     & 1.5492 & 1.2447 & 0.8404 & 1.7680 & 36.55 & 26.37       \TBstrut  \\
SE-ResNet-8  & 1.5516 & 1.2456 & 0.8512 & 1.8292 & 37.34 & 26.71       \TBstrut  \\ \hline
\end{tabular}
\end{adjustbox}
\end{table}

\subsection{Analysis of Model Complexity}
Here we discuss the differences in model complexity in our results and associate them to their relative performance to assess which instances display the best trade-off.

Starting with ResNet-3, we append a new block twice as deep as the previous one, resulting in ResNet-4 with $2.05$ million trainable parameters, and we repeat that for ResNet-5, reaching $7.60$ million. If we were to proceed with this pattern, the next model would reach $29.19$ million parameters, which is unnecessarily large. For reference, ResNet-50 \cite{He_2016_CVPR} itself has $25$ million. The same reasoning applies to the VGG architectures. In order to avoid a brute force procedure, the approach for expanding the models was to replicate the middle layer first and then alternate between a prior and posterior layer, moving towards both ends of the convolutional pipeline. That process is followed from ResNet/VGG-5 up to ResNet/VGG-10. 

Table \ref{complexity_table} summarizes all architectures in terms of the number of blocks and their respective depth values, enlisting the number of trainable parameters for each case. VGG-6 has the lowest error among the VGGs, displaying the best complexity trade-off since it only has $6.73$ million parameters. As for the ResNets, even though we observed ResNet-10 as the model with the best metrics, the improvement obtained is incremental for the complexity added, which doesn't justify using it. Also, the results for ResNet-9 are not consistently better than its predecessor.
Such remarks favor ResNet-8 as the most appropriate model depth. Finally, Table \ref{se_complexity_table} compares VGG-6 and ResNet-8 with their SE-augmented versions. 
Although the complexity added by the SE blocks is indeed minimal, mainly due to the low number of parameters from their two FC layers, the metrics still favored the original models.

\begin{table}[!h]
\caption{Model complexity comparison: number of trainable parameters in VGGs and ResNets}
\label{complexity_table}
\centering
\begin{adjustbox}{width=0.8\columnwidth}
\begin{tabular}{l|l|l|l}
\cline{3-4}
\multicolumn{2}{c}{}                    & \multicolumn{2}{c}{\# Trainable Parameters} \TBstrut  \\ \hline
X   & \multirow{1}{*}{Depth Summary}    & VGG-X & ResNet-X\\
\hline
\hline
3     & $[64^{\ },128^{\ }, 256^{\ }] $ & 397k & 580k      \TBstrut  \\ 
4     & $[64^{\ },128^{\ }, 256^{\ }, 512^{\ }] $ & 1.59M & 2.05M      \TBstrut  \\ 
5     & $[64^{\ },128^{\ }, 256^{\ }, 512^{\ }, 1024^{\ }]$ & 6.34M & 7.60M     \TBstrut  \\ 
6     & $[64^{\ },128^{\ }, 256^{2} , 512^{\ }, 1024^{\ }]$ & 6.73M & 7.99M     \TBstrut  \\ 
7     & $[64^{\ },128^{\ }, 256^{2} , 512^{2} , 1024^{\ }]$ & 8.30M & 9.57M     \TBstrut  \\ 
8     & $[64^{\ },128^{2} , 256^{2} , 512^{2} , 1024^{\ }]$ & 8.40M & 9.66M     \TBstrut  \\ 
9     & $[64^{2} ,128^{2} , 256^{2} , 512^{2} , 1024^{\ }]$ & 8.43M & 9.69M     \TBstrut  \\ 
10    & $[64^{2} ,128^{2} , 256^{2} , 512^{2} , 1024^{2}]$ & 14.72M & 15.98M    \TBstrut  \\ \hline
\end{tabular}
\end{adjustbox}
\end{table}

\begin{table}[!h]
\caption{Model complexity comparison: number of trainable parameters in SE-augmented models}
\label{se_complexity_table}
\scriptsize
\centering
\begin{adjustbox}{width=.55\columnwidth}
\begin{tabular}{l|c}
\hline
& \# Trainable Parameters \TBstrut  \\
\hline
\hline
VGG-6       & 6,730,907         \TBstrut  \\
ResNet-8    & 9,664,923         \TBstrut  \\
SE-VGG-6    & 6,916,071         \TBstrut  \\ 
SE-ResNet-8 & 9,885,583         \TBstrut  \\ \hline
\end{tabular}
\end{adjustbox}
\end{table}

\subsection{Performance Comparison}

In the following, we compare the results obtained by the benchmarks against our best deep learning models. {\reviewcolor All models were subjected to the same experiments, including the same data, number of samples, and data splits, so that the differences in performance are a reflection of their different capacities.} Table \ref{results_table} shows the performance metrics over the validation set. Among the machine learning benchmarks, the MLP-2 neural network works best, followed by the random forest. As for the related work benchmarks, ST-NN-TimeDNN \cite{2017arXiv171004350J} has relatively low performance and it is not a good fit for our dataset complexity. DNN \cite{deep_net_travel_2019}, on the other hand, is deep enough to outperform the classical benchmarks. Finally, SB-TTE \cite{wang_simple} works remarkably well despite the simplicity of the model. Lastly, ResNet-8 has the best results, followed by VGG-6 and SB-TTE. Although we elect one model as the best solution for our problem, we still provide a series of alternatives for solving it.

\begin{table}[!h]
\caption{Results Comparison}
\label{results_table}
\centering
\begin{adjustbox}{width=\columnwidth}
\begin{tabular}{l|l|l|l|l|l|l}
\hline
& MSE & RMSE & MAE & $\text{EW}_{90\%}$ & MARE & MAPE \\
\hline
\hline
SB-TTE \cite{wang_simple}                   & 2.0918 & 1.4463 & 0.9728 & 2.0698 & 39.34 & 30.51       \TBstrut  \\
ST-NN-TimeDNN \cite{2017arXiv171004350J}    & 2.3629 & 1.5372 & 1.1891 & 2.2197 & 56.87 & 37.31       \TBstrut  \\
DNN \cite{deep_net_travel_2019}             & 2.2649 & 1.5050 & 1.1377 & 2.2032 & 50.34 & 35.69       \TBstrut  \\ \hline
Random Forest          & 2.4432 & 1.5631 & 1.2081 & 2.2573 & 57.76 & 37.90     \TBstrut  \\ 
MLP-1        & 2.4506 & 1.5655 & 1.2147 & 2.2749 & 58.07 & 38.11     \TBstrut  \\
MLP-2        & 2.2940 & 1.5146 & 1.1587 & 2.2017 & 53.88 & 36.36     \TBstrut  \\ \hline
VGG-6       & 1.6979 & 1.3030 & 0.8867 & 1.9160 & 39.36 & 27.82         \TBstrut  \\ 
SE-VGG-6    & 1.7048 & 1.3057 & 0.9003 & 1.8814 & 39.86 & 28.25         \TBstrut  \\
ResNet-8    & 1.5492 & 1.2447 & 0.8404 & 1.7680 & 36.55 & 26.37         \TBstrut  \\
SE-ResNet-8 & 1.5516 & 1.2456 & 0.8512 & 1.8292 & 37.34 & 26.71         \TBstrut  \\ \hline
\end{tabular}
\end{adjustbox}
\end{table}

{\reviewcolor Table \ref{results_table} illustrates that our models (VGG-6, ResNet-8, and their SE-augmented versions) outperform the benchmarks by a significant margin. For example, while the best benchmark (SB-TTE) had an MARE of $30.51\%$, our best model (ResNet-8) showed $26.71\%$. Similarly for $\text{EW}_{90\%}$, the window of error reduced by $36$ minutes between these two models. Additionally, we verified the statistical significance of our results by means of Paired Sample T-tests performed between our best-performing model (ResNet-8) and each related work method in Table \ref{results_table}, for which we obtained p-values $p < 0.01$. However, the results obtained by the deep learning techniques implemented through our work are generally within close proximity of each other. This is due to the fact that the 4 chosen deep neural networks are effectively capable of learning the complex problem space almost similar to one another.}

{\reviewcolor Table \ref{results_table_runtime} compares our proposed model against the related work methods with respect to runtime, measured as the time taken for each model to run over the validation set. We observe that while our model took longer than ST-NN-TimeDNN \cite{2017arXiv171004350J} and DNN \cite{deep_net_travel_2019} due to its deeper architecture, the runtime is still reasonable, especially considering the relative gain in performance. Such an improvement could have considerable impacts on enhancing user experience. Besides, considering all challenges involved in such an OD problem, an automated solution that requires no route modeling (route-free) and excludes humans from feature designing (end-to-end) justifies leveraging deep learning towards tackling this problem, despite the need for additional complexity and computing resources. Lastly, we should point out that the reason why SB-TTE \cite{wang_simple} took orders of magnitude longer than the other methods is that for each input sample it searches through all the neighboring samples in the training set, resulting in extremely long processing times.}

\begin{table}[!h]
\caption{Runtime Comparison}
\label{results_table_runtime}
\centering
\begin{adjustbox}{width=.7\columnwidth}
\begin{tabular}{l|l|l|l}
\hline
& MARE & MAPE & Runtime \\
\hline
\hline
ST-NN-TimeDNN \cite{2017arXiv171004350J} & 56.87 & 37.31 & 6.48 s     \TBstrut  \\
DNN \cite{deep_net_travel_2019} & 50.34 & 35.69 & 20.87 s    \TBstrut  \\
SB-TTE \cite{wang_simple} & 39.34 & 30.51 & 9:15 h       \TBstrut  \\ \hline
ResNet-8 & 36.55 & 26.37 & 208.21 s   \TBstrut  \\ \hline
\end{tabular}
\end{adjustbox}
\end{table}

\subsection{Error Analysis}

We discuss the error distribution across the data and how that relates to the behavior of ResNet-8. For visualization purposes, we concentrate on the MAPE and MARE metrics.

\begin{figure}[!t]
\centering
\includegraphics[width=0.7\columnwidth]{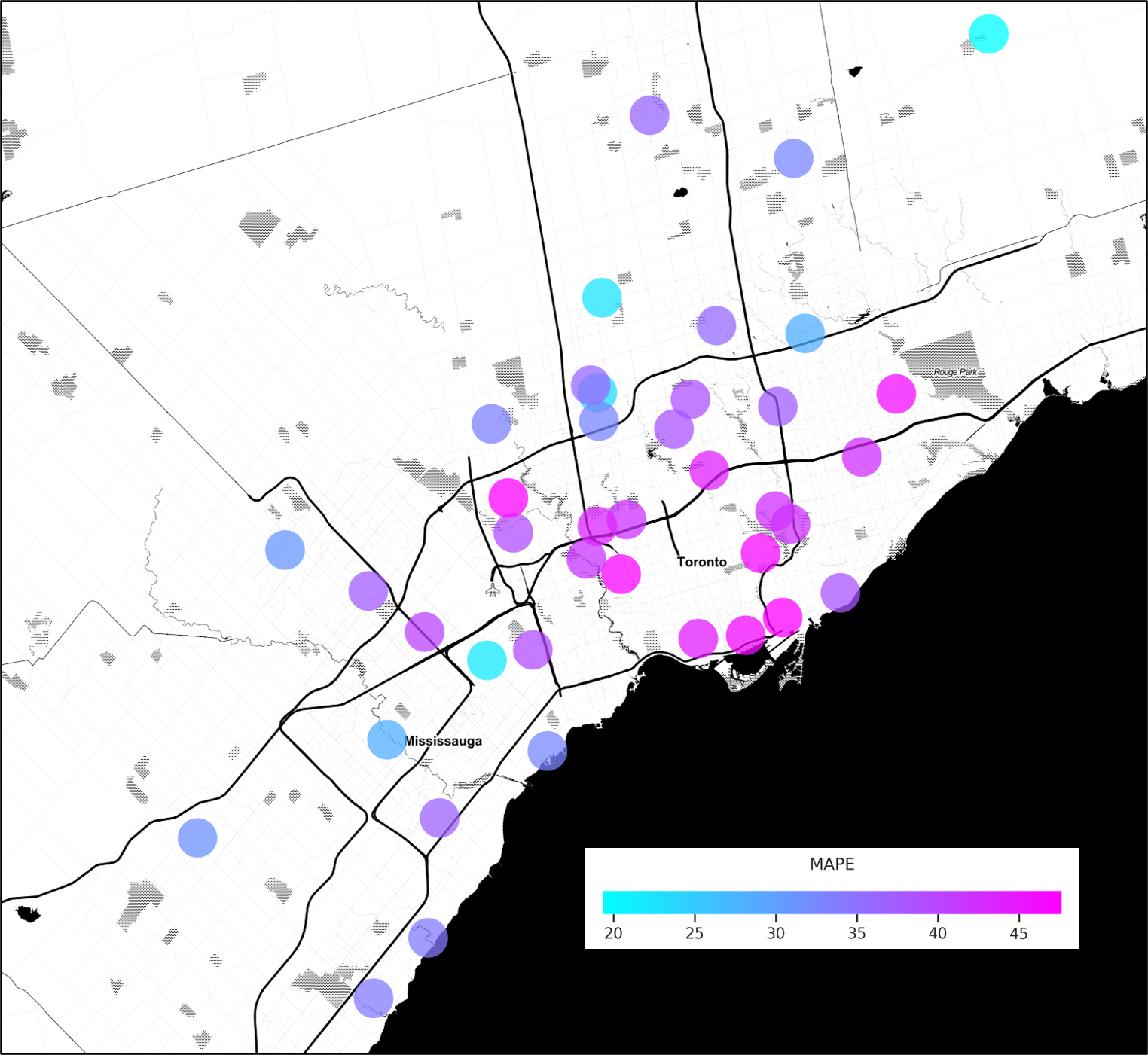}
\caption{Geographical distribution of MAPE across the $40$ busiest depots. A color map is used to encode the MAPE, ranging from light blue (low error), usually in the outskirts of GTA, to pink (high error) in (Downtown) Toronto.}
\label{error_analysis_vs_map}
\end{figure}

\subsubsection{Variations across depots}

we have discussed the imbalance between the number of origins and destinations in a parcel delivery problem, and the assumption of different spatio-temporal patterns for each depot. Figure \ref{error_analysis_vs_map} illustrates the spatial distribution of the errors and depots, overlaid onto a map of the GTA. Circles of equal radius are drawn at each location and a color map is defined according to the MAPE. As shown, bluer circles with lower MAPE tend to be found on the outskirts of GTA, possibly due to easier traffic or to a greater portion of routes being covered on highways. Conversely, depots located near Downtown Toronto (pink circles) display much higher MAPE.

\begin{figure}[!t]
\centering
\includegraphics[width=0.9\columnwidth]{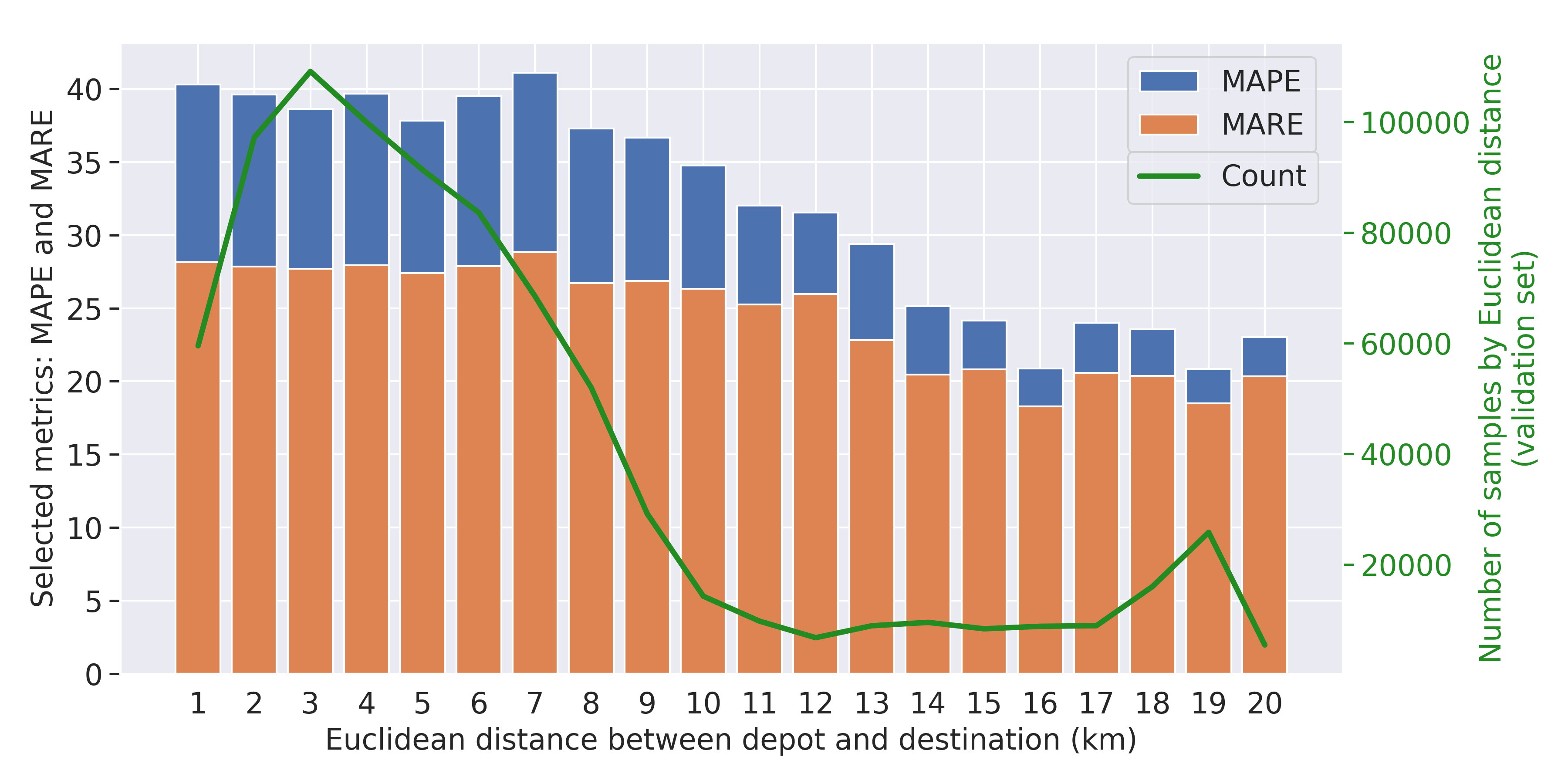}
\caption{Distribution of MAPE and MARE across the euclidean distance between the depot and the delivery destination. A green curve overlaid shows the number of samples per depot.}
\label{error_analysis_vecdist}
\end{figure}

\subsubsection{Variations across OD distances}

we now analyze the error variations against the direct distance between origin and destination. Figure \ref{error_analysis_vecdist} displays the error distribution versus the euclidean distance binned into integer kilometers. The performance is nearly the same up to $7$ km, where the data concentrates, indicating that, within that interval, this feature has no direct relation to the prediction ability.
Interestingly, the metrics do improve for longer distances, suggesting they are more predictable despite their low sample density (smaller training set). This may be due to the fact that longer distances often entail traveling through highways which are less susceptible to traffic and unforeseen driving circumstances.

\begin{figure}[!t]
\centering
\includegraphics[width=0.9\columnwidth]{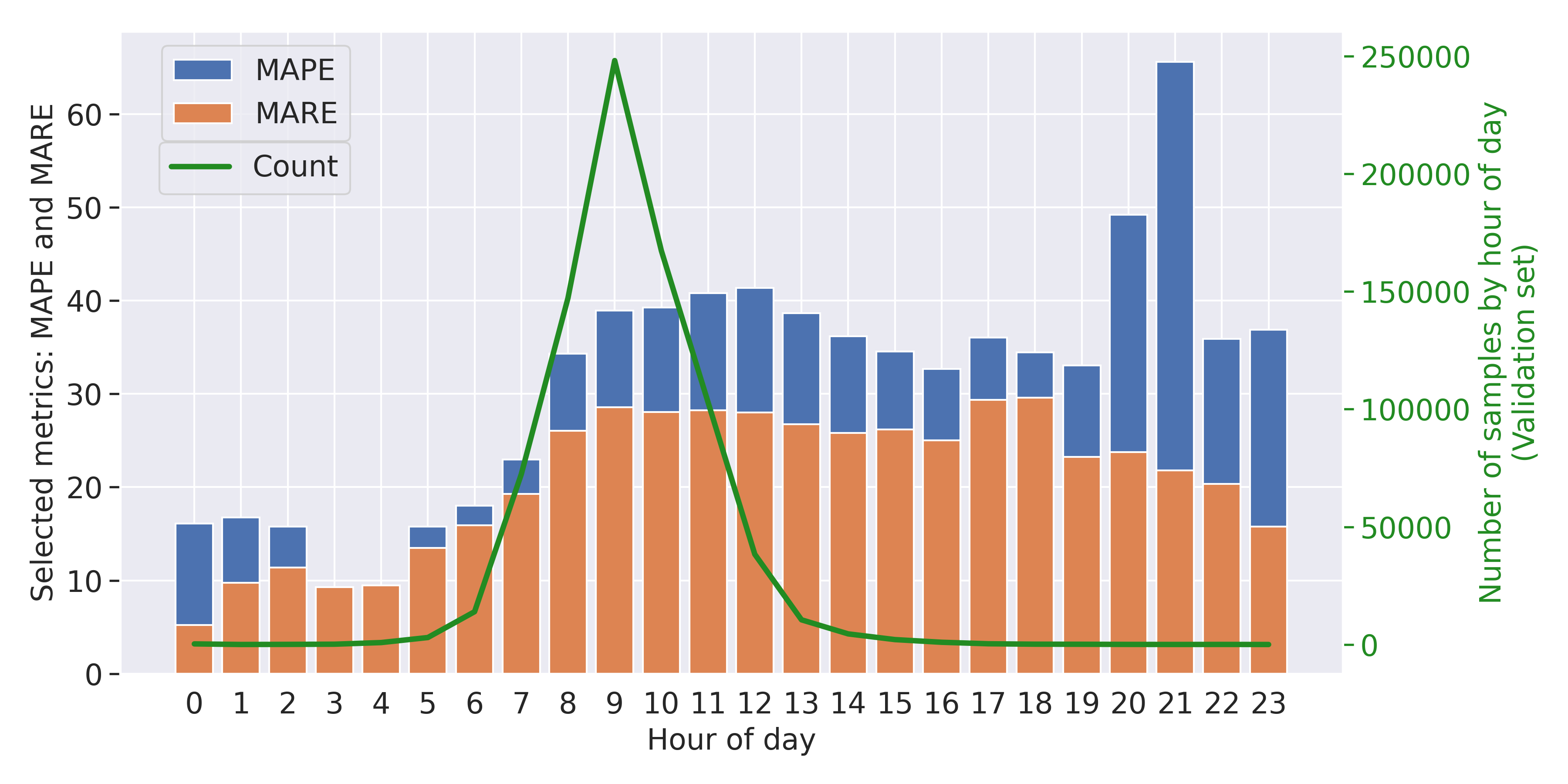}
\caption{Distribution of MAPE and MARE across the out-for-delivery hour-of-day. A green curve shows the distribution of samples throughout the day.
}
\label{error_analysis_3}
\end{figure}

\subsubsection{Variations across temporal features}

we analyze the distribution of error across temporal features, since traffic, and therefore predictability, are intrinsically time-variant phenomena. Figure \ref{error_analysis_3} shows the distribution of the MAPE and MARE, calculated over the validation set, versus the hour-of-day when the parcel was sent out for delivery. A green curve showing the distribution of samples tells that the majority of parcels are sent out in the morning, peaking around at $9$ a.m. The number of samples drastically decreases before $7$ a.m. and after noon, which could relate to the protocols and policies of Canada Post for last-mile delivery. Finally, the metrics show smaller errors for earlier hours, indicating that deliveries sent out earlier are likely more predictable, while the model performs approximately the same from $9$ a.m. to noon.

\begin{figure}[!t]
\centering
\includegraphics[width=0.9\columnwidth]{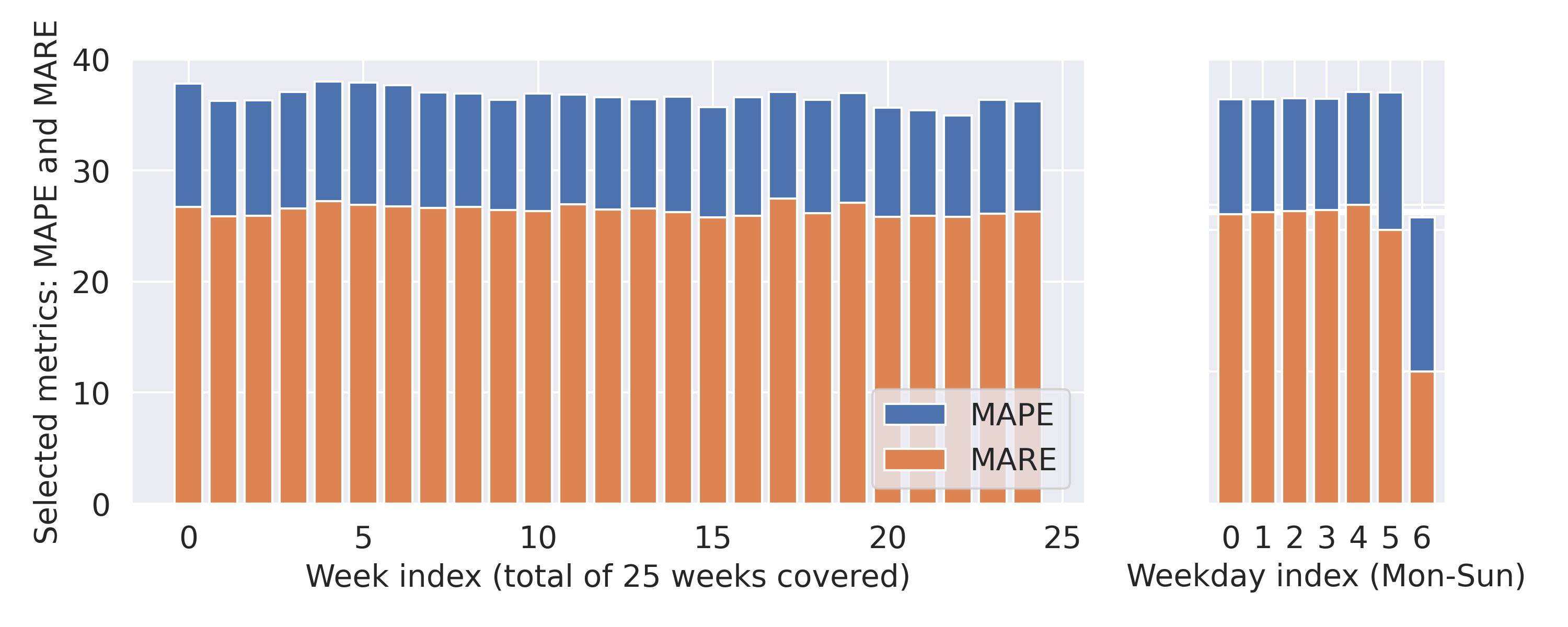}
\caption{Distribution of the MAPE and MARE across the $25$ weeks (left), and $7$ days of the week (right).
}
\label{error_analysis_vs_weekday}
\end{figure}

Moreover, Figure \ref{error_analysis_vs_weekday} presents error distributions against entire weeks, for a total of $25$ weeks covered in the dataset (6 months), and against the days of the week. Both metrics have shown no relevant variation across different weeks, so the model performs roughly the same every week, which is unexpected as it would be reasonable to anticipate higher errors in winter and lower errors in spring. 
As for days of the week, the error shows a slight decrease for Saturdays and a big improvement for Sundays, which could be due to the fact that weekends generally have better flowing traffic.

\subsubsection{Variations across targets}
\begin{figure}[!t]
\centering
\includegraphics[width=\columnwidth]{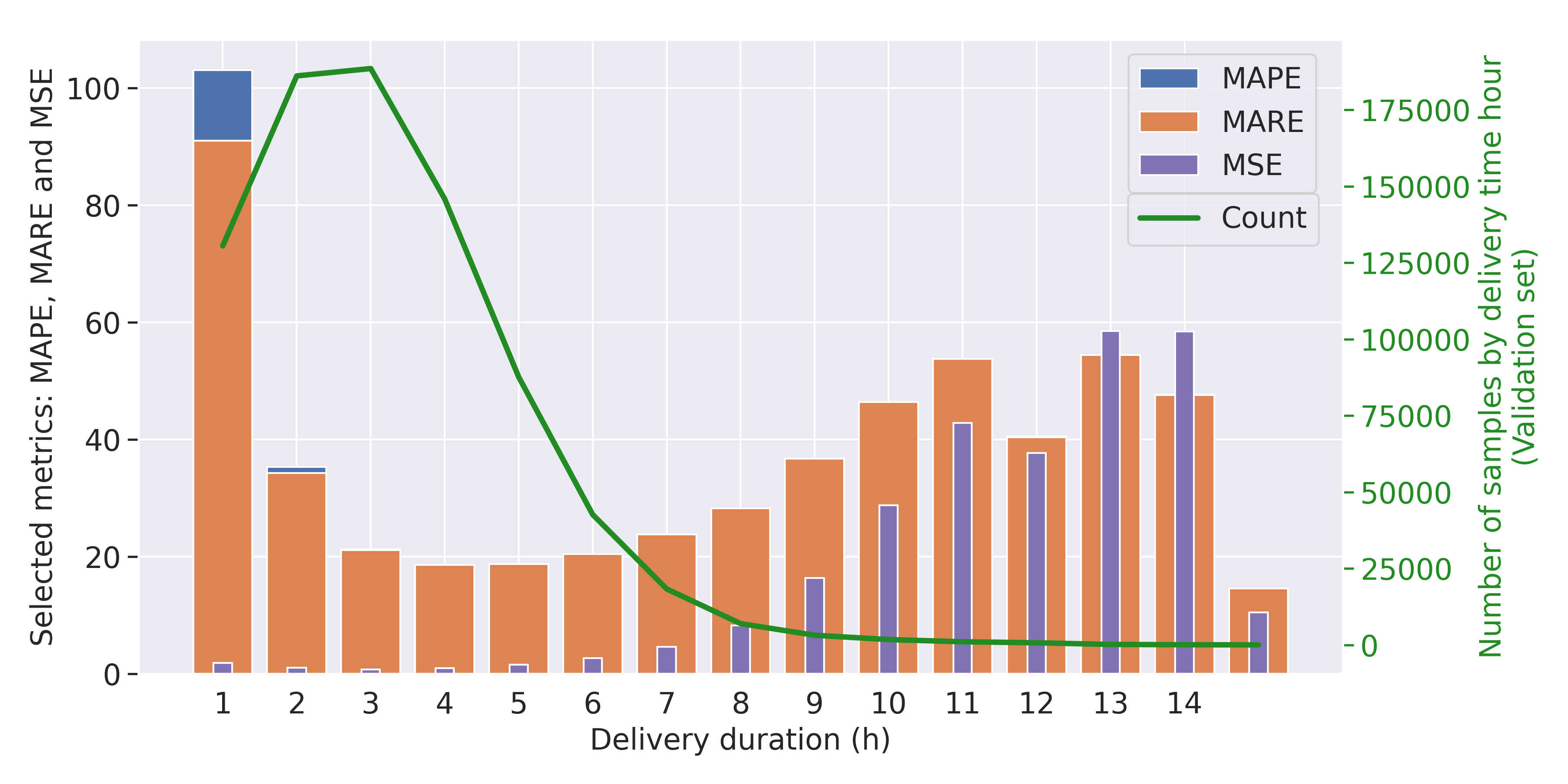}
\caption{Distribution of the MAPE, MARE, and the MSE across binned hours for the prediction target (delivery duration). A green curve shows the data distribution across different delivery durations.}
\label{error_analysis_vs_target}
\end{figure}

finally, we observe the performance for different delivery durations in Figure \ref{error_analysis_vs_target}. Since this is the model target, the loss (MSE) is displayed in addition to the metrics. We show in green the number of samples per hour. Considerably low MSE is obtained for deliveries taking up to $7$ hours, while for longer ones we observe that the loss increases with the duration. That could hint on the predictability of short versus long deliveries, but the MSE definition (\ref{equation_mse}) also plays a role. Based on squared absolute durations, it is directly affected by the target: while a $20\%$ error over a 12-hour delivery gives an MSE of $5.76$, the same $20\%$ error over a 1-hour delivery returns an MSE of $0.04$. As for the metrics, we observe an increase with duration, except in the first few bins, which we associate to MAPE/MARE being sensitive to errors for small targets, but also to the fact that deliveries that short might be abnormal and eventually need to be reviewed in the data. Interestingly, by grouping samples according to delivery time, the MAPE and MARE are roughly the same at each bin, which can be easily verified at the limit of making $y_i$ constant in Equations \ref{metrics_mape} and \ref{metrics_mare}.

The analysis of error with respect to both spatial and temporal dimensions helps in better understanding the model behavior, providing insightful interpretations on the predictability of parcel deliveries. Moreover, it confirms prior assumptions on the relevance of the selected features for prediction purposes.

\subsection{Visualizing Learned Representations}
As an attempt to better understand the model behavior, we present visualizations on the model representation of data across some of its features. Essentially, we project the learned representations onto a $2D$ space and observe how the model deals with some of the features in the data.

\begin{figure}[!t]
\centering
\includegraphics[width=.9\columnwidth]{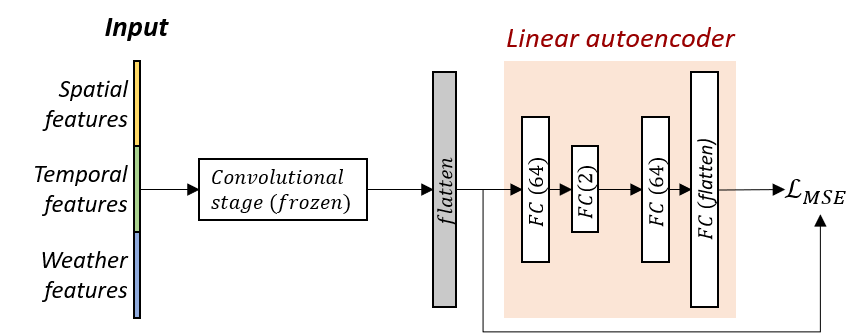}
\caption{An undercomplete autoencoder with linear activations replaces the FC layers of the model, projecting the learned representation onto a $2D$ space. The original model is frozen and the autoencoder training is unsupervised.}
\label{lin_autoencoder_pca}
\end{figure}

Since our model maps the data onto a very non-linear high-dimensional space, our approach to reducing dimensionality is to utilize principal component analysis (PCA) \cite{Abdi_Williams_2010}. We move away from the direct use of PCA since solving the eigenvalue decomposition for our dataset becomes unfeasible. In order to access the data’s principal components, we utilize an autoencoder, a neural network that is essentially trained to copy its input to its output \cite{Goodfellow-et-al-2016}. The network encodes the input data onto a latent representation and then decodes it back to the original space. An \textit{undercomplete} autoencoder, more specifically, has its latent layer with a smaller dimension than the input, being forced to discover the most informative dimensions in the data. Finally, an undercomplete autoencoder with linear activations optimized for MSE spans the same subspace as PCA \cite{Goodfellow-et-al-2016}. Therefore, we freeze our model, so their weights are no longer updated, and replace the FC layers by an autoencoder, which reduces the dimension down to $2$. The training is unsupervised, optimizing for the reconstruction loss (MSE). The autoencoder is summarized in Figure \ref{lin_autoencoder_pca}.

\begin{figure}[!t]
\centering
\includegraphics[width=\columnwidth]{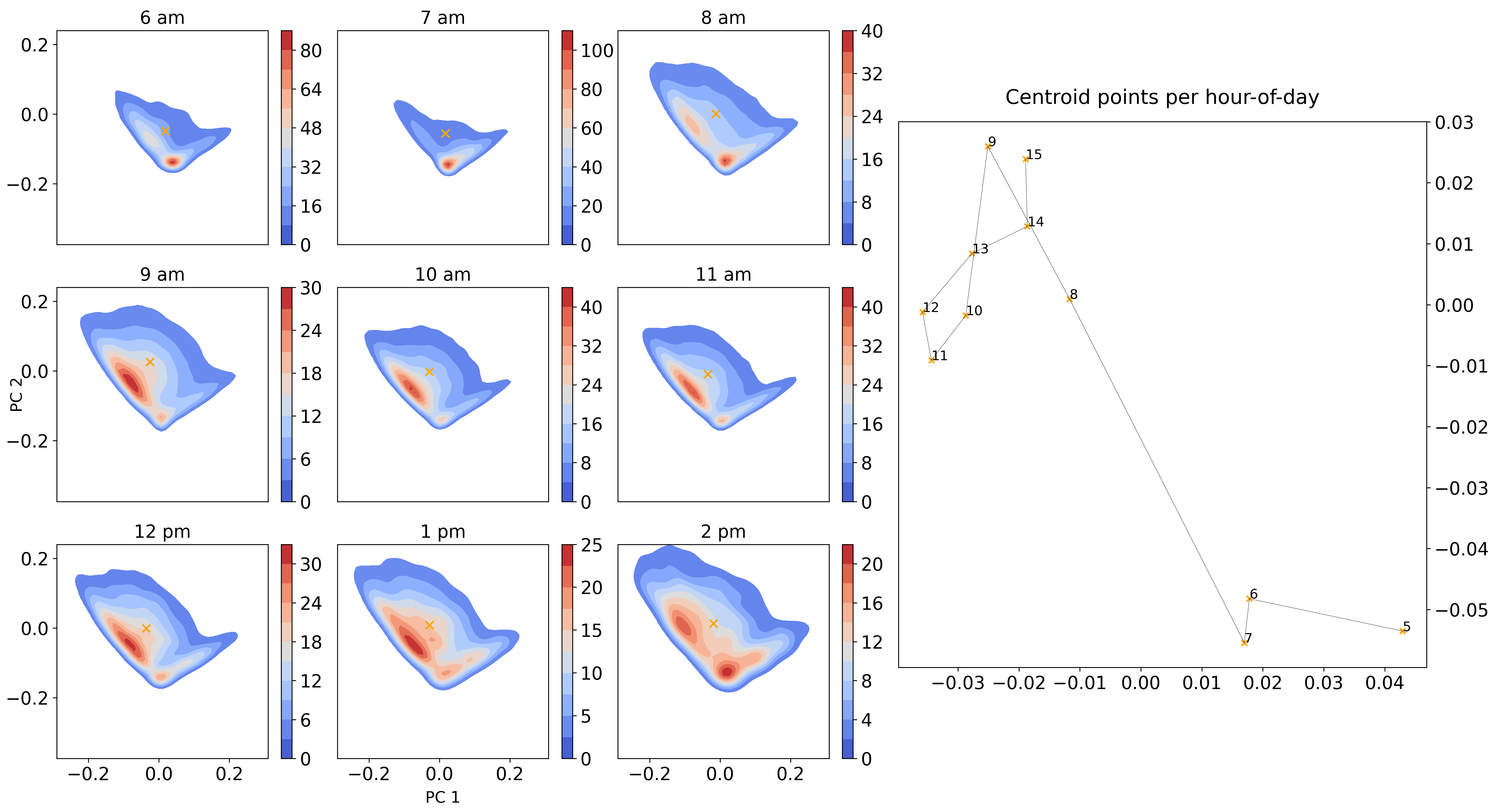}
\vspace{-0.5cm}\caption{The two principal components from the learned deep representations of the data are plotted against each other, from 6 a.m. to 2 p.m. On the right side, we track the position of the distribution centroids.}
\label{proj_2d_hour}
\end{figure}

Once trained, the encoded $2D$ representations of the data are collected. Figure \ref{proj_2d_hour} shows the distribution of the $2$ units at encoder output. Given their equivalence to PCA, we refer to them as principal components for clarity. The representations are shown by hour-of-day, from the 6 a.m. to 2 p.m period when most deliveries take place (see Figure \ref{error_analysis_3}). The distribution shifts and deforms as time goes by, showing a similar representation for $6-7$ a.m., as well as for $10-12$ a.m. Also, the distribution centroid is tracked on the right plot, showing major distribution shifts between $7$, $8$, and $9$ a.m. Essentially, this visualization hints at how the model prioritizes differentiating data from the busiest hours, assigning similar representations to those in scarcer bins. Still, interpreting the learned representations of a deep neural network is not trivial and, given that the variations on distributions over different hours are rather smooth, assessing the model behavior based on that discrimination is open to interpretation. 

\begin{figure}[!t]
\centering
\includegraphics[width=.8\columnwidth]{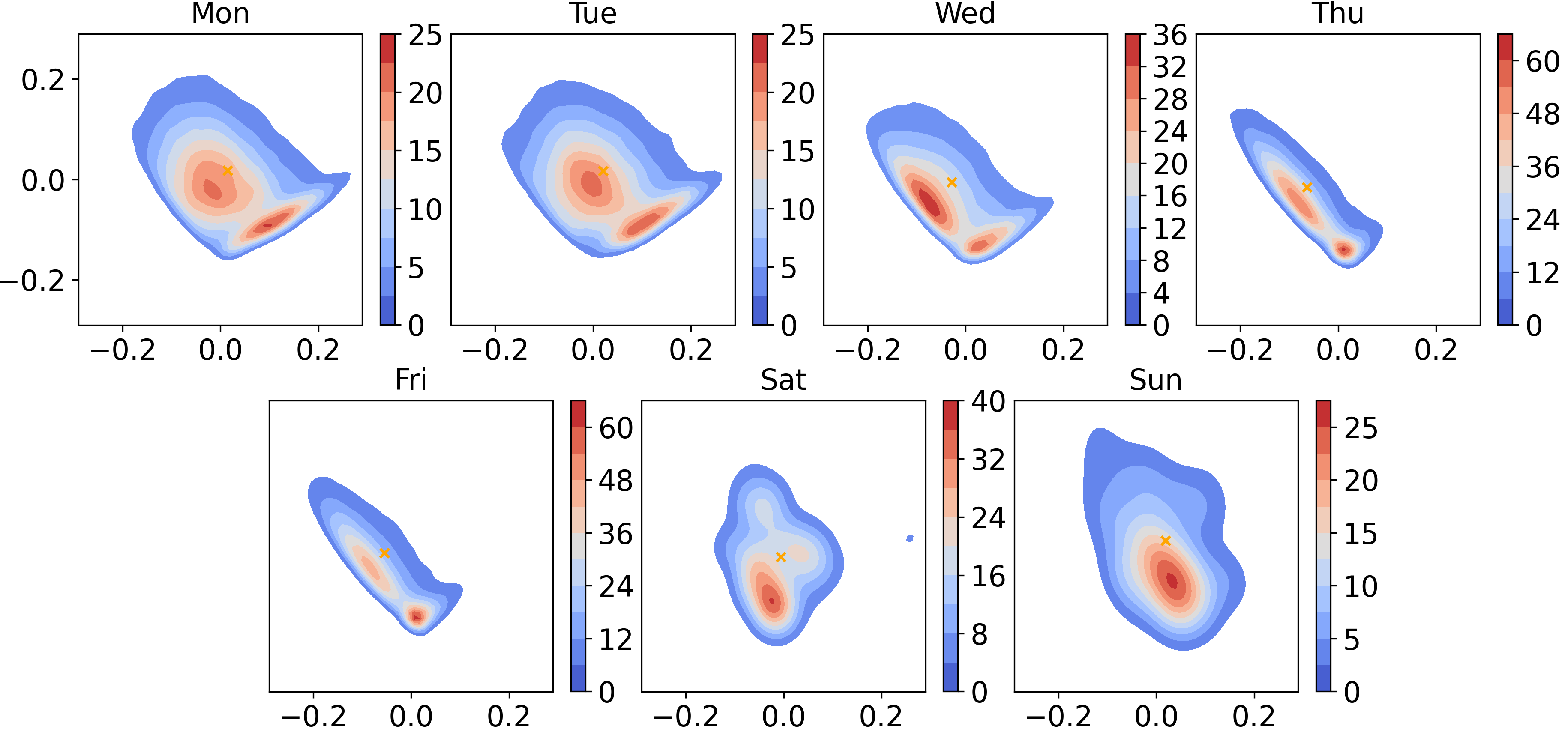}
\caption{The two principal components from the learned deep representations of the data are plotted against each other, across the five days of the week.}
\label{proj_2d_day}
\end{figure}

In Figure \ref{proj_2d_day} we analyze the projections with respect to the day-of-week. In principle, the model has learned similar representations for Mondays and Tuesdays, as well as for Thursdays and Fridays, indicating the importance of that feature to predicting delivery times. Interestingly, by representing the data as such, the model shows roughly the same performance for each weekday, as discussed in Figure \ref{error_analysis_vs_weekday}.

{\reviewcolor
\subsection{Limitations}
Although the route-free OD formulation could be interpreted as one that introduces limitations to performance, its premise is actually to move away from the reliance on uncertain error-prone routes. Still, one of the limitations in this work concerns the data: relying only on OD pairs of coordinates and timestamps would itself establish a lower bound of performance, and there are other aspects of the life cycle of parcels that could benefit performance while maintaining an OD-based formulation. Another limitation relates to modeling delivery times only at the moment when the parcel leaves the depot, with no further updates throughout the delivery. While our system manages to provide accurate predictions, it would certainly benefit from the ability to offer live updates to the customer, especially in the case of unexpected events.}

\section{Conclusions}

Our work focused on last-mile parcel delivery time estimation. We used a real-world difficult large-scale dataset of parcels delivered in the GTA during the first 6 months of $2017$, provided by Canada Post, the main postal operator in Canada. We present our solution as a smart city application under the IoT paradigm and discuss how a cloud-based architecture could render such a system feasible for real-world usage. With an OD-TTE formulation, we only rely on the \textit{out-for-delivery} and \textit{delivered} parcel scans, taking into account spatio-temporal information as well as weather data.
We explore several deep CNNs, and compare them against multiple benchmarks, from classical machine learning models to OD-based works from the literature. We demonstrate that a ResNet with 8 residual convolutional blocks achieves the best results, outperforming the other methods while providing a good performance-complexity balance. 
Further, we seek a better understanding of the model by thoroughly analyzing the errors across different features, and by visualizing the learned deep features through dimensionality reduction, which has led to interesting remarks on data predictability.
Our work delivers an end-to-end neural pipeline tailored to leverage parcel OD information as well as weather data to estimate accurate predictions.

A possible future direction would be expanding the data and pivot the approach to the driver’s perspective, knowing, for example, which driver made the delivery and their delivery history, and how many parcels were being delivered on the same day. Should the order of deliveries be available, Recurrent Neural Networks (RNN) could be used for time-series modeling of parcel sequences, allowing for more frequent updates to the user about the remaining parcels in the truck. Moreover, auxiliary inputs like traffic measurements or a map of potential routes could help to enhance performance.

\section*{Acknowledgment}
The authors would like to thank Innovapost Inc., Natural Sciences and Engineering Research Council of Canada (NSERC),
and Ontario Centres of Excellence (OCE) for funding this work.

\ifCLASSOPTIONcaptionsoff
  \newpage
\fi

\bibliography{IEEEabrv, refs.bib}

\end{document}